\documentclass[11pt,a4paper]{article}
\usepackage{amssymb,amsfonts,cite,psfrag,latexsym}
\usepackage[dvips]{graphicx}
\advance \topmargin by -\headheight
\advance \topmargin by -\headsep
\newcommand{\vs}[1]{\rule[- #1 mm]{0mm}{#1 mm}} 
\newcommand{\hs}[1]{\hspace{#1 mm}}

\newcommand{\lbl}[1]{\label{eq:#1}}
\newcommand{\rf}[1]{(\ref{eq:#1})}
\newcommand{\nn}{\nonumber}
\newtheorem{lem}{Lemma}[section]
\newtheorem{Thm}{Theorem}[section]

\newcommand{\be}{\vs{2}\begin{equation}}
\newcommand{\ee}{\vs{2}\end{equation}}
\newcommand{\cbe}{\vs{2}\[}
\newcommand{\cee}{\vs{2}\]}
\newcommand{\bea}{\begin{eqnarray}}
\newcommand{\eea}{\end{eqnarray}}
\newcommand{\nnbea}{\begin{eqnarray*}}
\newcommand{\nnena}{\end{eqnarray*}}
\newcommand{\leqa}{\lefteqn}

\newcommand{\dl}{\delta}
\newcommand{\Dl}{\Delta}
\newcommand{\eps}{\varepsilon}
\newcommand{\Ga}{\Gamma}
\newcommand{\ga}{\gamma}

\newcommand{\la}{\lambda}
\newcommand{\Om}{\Omega}
\newcommand{\om}{\omega}
\newcommand{\sg}{\sigma}
\newcommand{\Th}{\Theta}

\newcommand{\vf}{\varphi}

\newcommand{\Cc}{\mathcal{C}}       
\newcommand{\D}{\mathcal{D}}
\newcommand{\E}{\mathcal{E}}
\newcommand{\F}{\mathcal{F}}
\newcommand{\G}{\mathcal{G}}
\renewcommand{\H}{\mathcal{H}}      

\renewcommand{\L}{\mathcal{L}}

\newcommand{\mono}{\mbox{\LARGE\itshape m}} 
\renewcommand{\P}{\mathcal{P}}

\newcommand{\aux}{{\mathchoice{\mbox{\sl w}}{\mbox{\sl w}}
                              {\mbox{\ssz\sl w}}{\mbox{\ssz\sl w}}}}

\newcommand{\C}{\mathbb{C}}         
\newcommand{\R}{\mathbb{R}}         
\newcommand{\Sf}{\mathbb{S}}        
\newcommand{\openone}{\mbox{l\hspace{-2mm}1}} 

\newcommand{\Sb}{\mathbf{S}}        

\newcommand{\opname}[1]{\mathop{\mathrm{#1}}\nolimits}
\newcommand{\Diff}{\opname{Diff}}   
\newcommand{\id}{\opname{id}}       
\newcommand{\Lie}{\opname{Lie}}     
\newcommand{\supp}{\opname{supp}}   

\newcommand{\eg}{e.g.}              

\newcommand{\hcm}{\hspace{1cm}}      

\def\0{{\vphantom{\dagger}}}    

\newcommand{\dps}{\displaystyle}
\newcommand{\nms}{\normalsize}

\newcommand{\ssz}{\scriptsize}

\newcommand{\del}{\partial}

\newcommand{\ox}{\otimes}

\newcommand{\zer}[1]{{\mathord{\mathop{#1}\limits^\circ}}}

\renewcommand{\.}{\cdot}          
\newcommand{\trl}{\triangle}

\newcommand{\blp}{\biggl(}
\newcommand{\brp}{\biggr)}

\newcommand{\lra}{\ \longrightarrow\ } 
\newcommand{\Lra}{\ \Longrightarrow\ } 

\newcommand{\thalf}{\tfrac{1}{2}}     

\newcommand{\Bar}[1]{\overline{#1}}

\newcommand{\sepword}[1]{\quad\mbox{#1}\quad} 

\newcommand{\set}[1]{\{\,#1\,\}}        

\newcommand{\wick}[1]{\mathopen:#1\mathclose:} 

\def\<#1,#2>{\langle#1\,,\,#2\rangle} 

\newcommand{\tfrac}[2]{{\textstyle{\frac{#1}{#2}}}}

\newcommand{\row}[3]{{#1}_{#2},\dots,{#1}_{#3}}  

\newcommand{\bx}{\bigotimes\ }
\newcommand{\1}{\mathbf{1}}
\newcommand{\cafard}{%
\parbox{10mm}{\begin{picture}(20,10)
\put(0,5){\line(-1,2){5}}
\put(0,5){\line(-1,-2){5}}
\put(10,5){\circle{20}}
\put(10,14){\circle{15}}
\put(10,-4){\circle{15}}
\put(20,5){\line(1,2){5}}
\put(20,5){\line(1,-2){5}}
\end{picture}}
}
\newcommand{\kg}{%
\parbox{19mm}{\begin{picture}(20,10)
\put(0,5){\line(-1,2){5}}
\put(0,5){\line(-1,-2){5}}
\put(10,5){\circle{20}}
\put(20,5){\line(2,1){30}}
\put(20,5){\line(2,-1){30}}
\qbezier(35,5)(35,10)(40,15)
\qbezier(35,5)(35,0)(40,-5)
\qbezier(45,5)(45,10)(40,15)
\qbezier(45,5)(45,0)(40,-5)
\end{picture}}
}
\newcommand{\p}{%
\parbox{15mm}{\begin{picture}(20,10)
\put(5,-10){\line(0,1){30}}
\put(5,20){\line(1,0){30}}
\put(35,20){\line(0,-1){30}}
\put(35,-10){\line(-1,0){30}}
\put(0,25){\line(1,-1){40}}
\put(0,-15){\line(1,1){15}}
\put(40,25){\line(-1,-1){15}}
\qbezier(15,0)(25,0)(25,10)
\end{picture}}
}
\newcommand{\ocateye}{
\parbox{11mm}{\begin{picture}(20,10)
\put(5,5){\line(-1,2){5}}
\put(5,5){\line(-1,-2){5}}
\put(15,5){\circle{20}}
\put(25,5){\line(1,2){5}}
\put(25,5){\line(1,-2){5}}
\qbezier(11,5)(11,10)(15,15)
\qbezier(19,5)(19,10)(15,15)
\qbezier(11,5)(11,0)(15,-5)
\qbezier(19,5)(19,0)(15,-5)
\end{picture}}
}

\newcommand{\pinterIV}{%
\parbox{16mm}{\begin{picture}(20,10)
\put(10,5){\line(-2,-1){10}}
\put(10,5){\line(-2,1){10}}
\put(10,5){\line(2,1){30}}
\put(10,5){\line(2,-1){30}}
\qbezier(25,5)(25,10)(30,15)
\qbezier(25,5)(25,0)(30,-5)
\qbezier(35,5)(35,10)(30,15)
\qbezier(35,5)(35,0)(30,-5)
\end{picture}}
}

\newcommand{\ppois}{%
\parbox{22mm}{\begin{picture}(20,10)
\put(0,5){\qbezier(0,-5)(20,25)(38,0)}
\put(0,5){\qbezier(0,5)(20,-25)(38,0)}
\put(36,5){\qbezier(2,0)(20,25)(40,-5)}
\put(36,5){\qbezier(2,0)(20,-25)(40,5)}
\end{picture}}
}

\newcommand{\sunset}{%
\parbox{15mm}{\begin{picture}(20,10)
\put(0,5){\line(1,0){10}}
\put(20,5){\circle{20}}
\put(10,5){\line(1,0){30}}
\end{picture}}
}
\newcommand{\iice}{%
\parbox{21mm}{\begin{picture}(20,10)
\put(0,5){\line(1,0){10}}
\put(20,5){\circle{20}}
\put(40.2,5){\circle{20}}
\put(30.2,5){\oval(40.5,40)[bl]}
\put(30.2,5){\oval(40,40)[br]}
\put(50.2,5){\line(1,0){10}}
\end{picture}}
}
\newcommand{\triple}{%
\parbox{28mm}{\begin{picture}(20,10)
\put(0,15){\line(1,0){80}}
\put(20,15){\circle{20}}
\put(60,15){\circle{20}}
\put(40,-25){\line(0,1){50}}
\put(40,-5){\circle{20}}
\end{picture}}
}
\newcommand{\sweet}{%
\parbox{10mm}{\begin{picture}(20,10)
\put(0,5){\line(-1,2){5}}
\put(0,5){\line(-1,-2){5}}
\put(10,5){\circle{20}}
\put(20,5){\line(1,2){5}}
\put(20,5){\line(1,-2){5}}
\end{picture}}
}
\newcommand{\dsweet}{%
\parbox{15mm}{\begin{picture}(20,10)
\put(0,5){\line(-1,2){5}}
\put(0,5){\line(-1,-2){5}}
\put(10,5){\circle{20}}
\put(30.2,5){\circle{20}}
\put(40.2,5){\line(1,2){5}}
\put(40.2,5){\line(1,-2){5}}
\end{picture}}
}
\newcommand{\tsweet}{%
\parbox{23mm}{\begin{picture}(20,10)
\put(0,5){\line(-1,2){5}}
\put(0,5){\line(-1,-2){5}}
\put(10,5){\circle{20}}
\put(30.2,5){\circle{20}}
\put(50.4,5){\circle{20}}
\put(60.2,5){\line(1,2){5}}
\put(60.2,5){\line(1,-2){5}}
\end{picture}}
}

\newcommand{\AIHPA}[1]{Ann. Inst. Henri Poincar\'e \textbf{A#1}}
\newcommand{\ATMP}[1]{Adv. Theor. Math. Phys. \textbf{#1}}

\newcommand{\CMP}[1]{Commun. Math. Phys. \textbf{#1}}
\newcommand{\EPJC}[1]{Eur. Phys. J. \textbf{C#1}}
\newcommand{\HPA}[1]{Helv. Phys. Acta \textbf{#1}}
\newcommand{\IJMPA}[1]{Int. J. of Mod. Phys. \textbf{A#1}}

\newcommand{\JMAA}[1]{J. Math. Anal. Appls. \textbf{#1}}
\newcommand{\JPA}[1]{J. Phys. \textbf{A#1}}

\newcommand{\NPB}[1]{Nucl. Phys. \textbf{B#1}}
\newcommand{\PLB}[1]{Phys. Lett. \textbf{B#1}}

\newcommand{\TMP}[1]{Theor. Math. Phys. \textbf{#1}}

\begin{document}

\font\fifteen=cmbx10 at 15pt
\font\twelve=cmbx10 at 12pt

\begin{titlepage}

\begin{center}

\indent

\vskip -1.5cm

\renewcommand{\thefootnote}{\fnsymbol{footnote}}

{\twelve Centre de Physique Th\'eorique,\footnote{
Unit\'e Propre de Recherche 7061}
CNRS Luminy, Case 907}

{\twelve F--13288 Marseille -- Cedex 9}

\vskip 2.cm

{\fifteen CONNES--KREIMER--EPSTEIN--GLASER\\[2mm] RENORMALIZATION}

\vskip 1.5cm

\setcounter{footnote}{0}
\renewcommand{\thefootnote}{\arabic{footnote}}

{\bf
Jos\'e M. GRACIA-BOND\'IA}$^{a,b,c}$
\hskip 0.2mm
and
{\bf Serge LAZZARINI} $^{c,}$\footnote{and also
Universit\'e de la M\'editerran\'ee, Aix-Marseille II,
e-mail\,: \texttt{sel@cpt.univ-mrs.fr}}
\\[6mm]
$^a$ {\it Institut f\"ur Physik, Theoretische Elementarteilchenphysik,\\
Universit\"at Mainz, D--55099 Mainz, GERMANY}\\[2mm]
$^b$ {\it Departamento de F\'{\i}sica Te\'orica I, Universidad
Complutense, Madrid 28040, SPAIN}\\[2mm]
$^c$ {\it Centre de Physique Th\'eorique, CNRS Luminy, Case 907,
F--13288 Marseille Cedex 9, FRANCE}
\end{center}

\vskip 1.5cm

\begin{abstract}
Causal perturbative renormalization within the recursive
Epstein--Glaser scheme involves extending, at each order, time-ordered
operator-valued distributions to coinciding points. This is achieved
by a generalized Taylor subtraction on test functions, which is
transposed to distributions. We show how the Epstein--Glaser recursive
construction can, by means of a slight modification of the Hopf
algebra of Feynman graphs, be recast in terms of the new
Connes--Kreimer algebraic setup for renormalization. This is
illustrated for $\phi^4_4$-theory.
\end{abstract}

\vskip 1.5cm

\noindent 1998 PACS Classification: 11.10.Gh Renormalization,
11.15.Bt General properties of perturbation theory.

\indent


\indent

\noindent{CPT--2000/P.4013, MZ--TH/00--30, FT/UCM--44--2000} \hfill
\today

\noindent{\tt hep-th/0006106}

\noindent web : \texttt{www.cpt.univ-mrs.fr}
\hfill{anonymous ftp : \texttt{cpt.univ-mrs.fr}}

\end{titlepage}
\setcounter{footnote}{0}

\section{Introduction}

\indent

A quarter of a century ago, the venerable art of bypassing infinities
seemed poised to reach the status of science. The ``state of the art''
at that time was summarized in the authoritative lectures at the 1975
Erice Majorana School, including outstanding treatments of dimensional
renormalization, BPHZ renormalization, the forest formula, the
Epstein--Glaser scheme and the BRS method for gauge models. That
School's book~\cite{VV} remains an indispensable classical in the
field. It is fair to say that, despite some technical and conceptual
advances, no major progress in the systematics of renormalization
theories took place again until recent dates.

Ever since Kreimer perceived the existence of a Hopf algebra lurking
behind the forest formula~\cite{DirkFirst}, however, the question of
encoding the systematics of renormalization in such a structure, and
of the practical advantages therein, brought the subject to the
forefront again.
The first attempt used an algebra of parenthesized words
or rooted trees to describe Feynman diagrams, and,
as quickly pointed out by Krajewski and Wulkenhaar~\cite{KW}, although able
in principle to deal with overlapping divergences
(see, for instance, the Appendix of~\cite{CoKrFirst}),
was ill adapted in practice to do so.
More recently, Connes and Kreimer~\cite{CoKrI,Connes,CoKrII}
were indeed able to show, using $\phi^3_6$ as an example, that
dimensional regularization of quantum field theories in momentum
space, and the renormalization group, are encoded in a commutative
Hopf algebra \textit{of Feynman graphs} $\H$ and an associated
Riemann--Hilbert problem.
Moreover, there exists a Hopf algebra map from the Connes--Moscovici
algebra $\H_{\rm CM}$~\cite{CM} of the group $\Diff'(\C)$ of nontrivial
diffeomorphisms of the complex plane to~$\H$.

After the achievement by Connes and Kreimer, a reformulation of
renormalized, perturbative quantum field theory in Hopf algebraic
terms is in the cards. Because $\H$ is commutative, its vast group of
characters $\G_\H$ contains the same information, and there is a
corresponding morphism of groups $\G_\H \to \Diff'(\C)$. Moreover, the
Lie algebra $\Lie(\G_\H)$ of infinitesimal characters, with bracket
operation given by diagram insertions~\cite{CoKrI}, and its enveloping
algebra, also contain in principle the same information. Therefore,
beyond its mathematical elegance and this tantalizing connection to
diffeomorphism-invariant geometry in noncommutative geometry, the Hopf
algebra approach holds the promise of a fruitful use of
$\G_\H$-invariance as a tool of the renormalized
theory~\cite{Connes,CoKrII,Rusos}.

The present paper wants to contribute to this program by considering a
different renormalization setup. Dimensional renormalization in effect
occupies a privileged rank among methods that use an intermediate
regularization. On the other hand, the most rigorous
regularization-free method is causal perturbation theory (CPT), i.e.,
the scheme devised by Epstein and Glaser
---see~\cite{EG73,Philippe}--- following the basic ideas of
Stueckelberg--Bogoliubov--Shirkov~\cite{Stu,BS80}, linking the problem
of ultraviolet divergences to questions in distribution theory. This
is not the place to go into the respective merits of
regularization-bound and regularization-free methods. Suffice it to
say that CPT used to have a reputation for difficulty; but nowadays
its tremendous power and flexibility are widely recognized. For
instance, differential renormalization~\cite{FJL}, especially in the
elegant form given to it in~\cite{ZavSmir}, has been shown in the
remarkable paper~\cite{Prange} as another instance of CPT. Also BPHZ
renormalization, in spite of its very different flavour, is just a
Fourier space translation of a particular case of
CPT~\cite{Prange,Scharf}.

Being a ``real'' space procedure, CPT is the natural candidate for
renormalization in curved backgrounds~\cite{BrFr99}. Also, the kinship
of CPT to the Segal--Shale--Stinespring formulation of quantum field
theory (i.e., the non-perturbative approach based on the metaplectic
and spin representations), spelled out by Bellissard in the
seventies~\cite{Bellissard}, then apparently forgotten, has
lately~\cite{Halley} been highlighted again.

The aim of this paper is to show that the Connes--Kreimer Hopf algebra
approach is consistent with the Epstein--Glaser renormalization
method. We also proceed mainly by way of example and, in particular,
we exhibit a Hopf algebra that encodes the renormalization of
$\phi^4_4$ scalar theory. Now, in a recent paper by
Pinter~\cite{Pinter}, in the same context of $\phi_4^4$, the
successive renormalization of superficial divergences and
subdivergences, buried in the abstract CPT procedure, was brought to
light for diagrams occurring in the expansion of the scattering matrix
up to third order. The present article extends Pinter's work to
combinatorially nontrivial cases, as well as pursuing the work
of~\cite{CoKrI,Connes,CoKrII}. The equivalence between the classical
recursive formulae for renormalization and the forest formula of
Zimmermann, which is shown to hold in the Epstein--Glaser procedure,
boils down to the result that the ``$C$-map'' and thus the ``$R$-map''
are in fact characters of our Hopf algebra.

\smallskip

We do not suppose that the reader is familiar with CPT. Accordingly,
some questions of principle relative to the conceptual and
mathematical status of the method are reviewed in Section~2 of the
paper. Sections~3 and~4 deal with the machinery of CPT, in the variant
proposed by Stora. Section~5 is concerned with the multiplicative
property of the counterterm map within causal perturbation theory. In
Section~6, we introduce an algebra structure $\H$ \`a la
Connes--Kreimer for the $\phi^4_4$ theory. The renormalization method
of Bogoliubov--Epstein--Glaser is then restated in the Hopf algebra
context. In Section~7, our treatment of CPT is further illustrated by
a few more examples. The conclusions follow.

We do suppose the reader to be familiar with the axioms of Hopf
algebra theory, as, for instance, in Chapter~3 of~\cite{Kassel} or in
Chapter~1 of~\cite{Polaris}, and with the basics of Bogoliubov's
$R$-map~\cite{BS80}.

\section{Motivating causal perturbative renormalization}
\label{motiv}

\indent

In CPT the emphasis is laid not so much on eliminating the infinities
apparent in the na\"{\i}ve bare Feynman amplitudes as on redefining
the objects of the theory in such a way that infinities are never met
in the first place. Thus, there can be some contention on whether CPT
is truly a renormalization theory in the ordinary sense. An example,
taken originally from~\cite{Prange}, will, we hope, illuminate the
matter.

In classical electrostatics, the electric potential $V$ and field
$\vec E$ are given as distributions on~$\R^3$. If there is a point
charge source $e$ at the origin, the solution of the Poisson equation
$\Dl V = -4\pi e\dl$ is given by the distribution $e/r$, an element of
$\D'(\R^3)$. The electric field is obtained thus:
\be
\vec E(\vec x) = - \vec\nabla V = \frac{e\,\vec x}{r^3} \in \D'(\R^3).
\ee
Now, suppose we are asked what is the total energy stored in the
field. The \textit{density} of energy would notionally be given by
$|\vec E|^2$. However, in general distributions cannot be squared, and
in particular it is clear that this product does not exist as an
element of $\D'(\R^3)$. One could take the attitude of saying that the
self-energy is infinite; or one could say, perhaps more accurately,
that the self-energy functional is undefined globally from physical
principles, in this field theory. On the other hand,
\be
|\vec E|^2(\vec x) = \frac{e^2}{r^4}
\ee
makes sense as an element of $\D'(\R^3\setminus\{0\})$. If we insist
on seeking a sensible answer to the question, we can try to make an
\textit{extension} of $r^{-4}$ to a distribution defined on the whole
of $\R^3$. This is certainly possible, albeit in a nonunique way.

A bit more generally, suppose that, as in the present case, $f$ is
defined as a regular function having only an algebraic singularity of
finite order at the origin. Then, a canonical way~\cite{EG73} of
making the extension of $f$ to a distribution defined on all of $\R^3$
is obtained by choosing an auxiliary function $\aux$ such that
$\aux(0) = 1$ and all its derivatives up to the order of singularity
$\om(f)$ of~$f$ vanish at the origin, and then defining the extension
$f_\aux$ by subtraction and transposition:
\be
\<f_\aux(\vec x), \vf(\vec x)>
 := \<f(\vec x), \vf_\aux(\vec x)>
 := \left< f(\vec x), \vf(\vec x) - \aux(\vec x)
    \smash{\sum_{|a|=0}^{\om(f)}} \frac{\del_a\vf(0)}{a!}\, x^a \right>,
\ee
where $a$ is a multi-index. This is CPT renormalization in a nutshell.

Naturally, the extension so defined depends on $\aux$, albeit in a
relatively weak way. In our case, $f = |\vec E|^2$ and because
$\om(f) = 1$ (i.e., it would be necessary to multiply
$|\vec E|^2(\vec x)$ as an integrand by $r^{1+\eps}$ to get a finite
integral), the nonuniqueness of the extension is expressed by the
possibility of adding distributions supported at the origin, of the
form $C^0\dl(\vec x) + C^i \del_i\dl(\vec x)$, to any particular
extension. The constants $C^0, C^i$ parametrize the indeterminacy in
the mathematical solution to the problem.

Actually, in the present case we can readily discard the three
constants $C^i$, for reasons of symmetry. Moreover, in view of the
good ``infrared behaviour'' of $|\vec E|^2$, it is possible to choose
for $\aux$ the natural candidate $\aux \equiv 1$. This gives for the
self-energy of the classical electron field:
\be
\<|\vec E|^2_1, 1> + C^0 = C^0.
\ee
According to this disposition, the energy is not so much infinite as
undetermined. One can then try to fix the value of $C^0$ by extra
physical information and/or assumptions. For instance, if we have
measured the mass of the point charge and we believe all this mass to
be of electrical origin, we must then believe that Nature selects
$C^0 = m$, in standard units.

Now, one can take the attitude that, even if no infrared problem is
present, it is more natural to make $\aux$ different from zero only in
a neighbourhood of the origin. Take, for instance,
$\aux(\vec x) = \Th(a^2 - |\vec x|^2)$, where $\Th$ denotes the
Heaviside step function ($\Th(t) = 1$ for $t \geq 0$ and $0$
otherwise). Then one reproduces the previous result \emph{directly} by
choosing $a = 4\pi e^2/m$.

There are several morals to the story, already: continuation of
distributions corresponds to renormalization; ``mass scales'' do
appear naturally in the process; sometimes, physical requirements
allow to lessen or sometimes completely eliminate the indeterminacy of
renormalization.

Now, CPT-like renormalization is not the only one available on the
market. In a more technical vein, let us explore some alternatives.
The extension of a function $f$ depending only on the radial
coordinate can always be reduced to a one-dimensional
problem~\cite{RicardoK}; note that in the mathematical literature, the
extension process is often called ``regularization'', a terminology
which for obvious reasons we cannot embrace here. In fact, on $\R^n$
with $\Sf$ the unit sphere, the equality
\be
\<f(r), \vf(\vec x)>_{\vec x} = \<\Th(r)r^{n-1}f(r), \tilde\vf(r)>_r
\ee
holds, where
\be
\tilde\vf(r) := \int_\Sf \vf(r\vec u) \,d\Om(\vec u)
\ee
is extended to all $r \in \R$ as an even function, as all its odd
order derivatives vanish at the origin. Our example is thus related to
possible definitions for $\Th(x)x^{-2}$. A popular one is obtained by
analytic continuation of $\Th(x)|x|^\la$; in this case the pole at
$\la = -2$ is removable. Another strategy consists in extracting the
``finite part'' of $\Th(x)x^{-2}$: knowing a priori that the integral
\be
G(\eps) := \int_\eps^\infty \frac{\vf(x)\,dx}{x^2}
\ee
is of the form $G(\eps) = G_0(\eps) + b\log\eps + c/\eps$, where $G_0$
has a finite limit as $\eps \downarrow 0$, the finite part is defined
to be this limit. Yet another idea is to define $\Th(x)x^{-2}$ as the
distributional second derivative of $-\Th(x)\log(x)$ ---this is
``differential renormalization''. The diligent reader will be able to
check that, for our case, the three procedures coincide. This little
miracle happens when extending $r^{-k}$ in $\R^n$ iff $k - n$ is odd.
The coincident extensions differ from an extension in the style of~(3)
with $\aux(\vec x) = \Th(a^2-|\vec x|^2)$, by terms that involve the
scale $a$ and the distributions $\dl$ and $\dl'$. An interesting
question, which apparently has not received much attention, is to find
a $\aux$ that reproduces a given extension.

To conclude, we already mentioned that the choice $\aux = 1$ does
nothing to improve the infrared behaviour of $f$. In order to obtain a
good infrared behaviour, however, it is not necessary to proceed to
sharp cutoffs; more precisely, it is enough that $\aux$ go to zero
rapidly in the Ces\`aro sense~\cite{Odysseus,RicardoSteve}. For
instance, the difference between BPHZ subtraction at zero momentum and
at $q \neq 0$ momentum in CPT corresponds essentially to the
replacement of~$1$ by an exponential function ---that decreases faster
than any inverse power of $x$ in the Ces\`aro sense--- as the
$\aux$-function in real space.

The example in this section differs less in substance than in
technical simplicity/complication from the treatments of the
divergences that crop up in quantum field theory.

\section{Surveying the Epstein--Glaser procedure}

\indent

Since we are concerned with the $\phi_4^4$ theory as a pedagogical
example, the discussion in this section will be centred on that
instance. We use the standard unit system ($\hbar = c = 1$) throughout
the paper. Let $M_4$ be the 4-dimensional spacetime endowed with the
Minkowski metric $(+,-,-,-)$. The following notation will be
repeatedly used.

\smallskip

\noindent
{\bf Notation}. According to the chosen signature, for any 4-vector
$\xi = (\xi^0,\vec{{}\xi})$ in any Lorentzian frame, with $\xi^0$
denoting the time component, the Minkowski metric is written
$\xi^2 = (\xi^0)^2 - \vec{{}\xi}{}^2$.

Let $\Bar V^-$ be the closure of the past light cone,
\be
\xi \in \Bar V^- \iff \xi^0 \leq 0 \mbox{ and } \xi^2 \geq 0
  \ (\mbox{or } \xi^0 \leq - |\vec{{}\xi}|).
\ee
For any pair of points $x,y\in M_4$, we shall write
\cbe
x \leq y \iff x - y \in  \Bar V{}^- ,
\cee
and for the complement,
\begin{equation}
x \gtrsim y \iff x - y \not\in \Bar V{}^-
 \iff x \in \complement \left(\{y\} + \Bar V{}^- \right)
 \iff x^0 - y^0 > 0 \mbox{ or } (x - y)^2 < 0,
\lbl{nopast}
\end{equation}
which in particular means that $x$ belongs to an open region in $M_4$.
Finally,
\cbe
x \sim y \iff x \gtrsim y \mbox{ \ and \ } y \gtrsim x
 \iff (x - y)^2 < 0,
\cee
meaning that $x$ and $y$ are spacelike-separated points. It is to be
also noted that
\cbe
x \leq y \mbox{ \ and \ } y \leq x  \Lra  x = y.
\cee

\subsection{The $\phi^4_4$ theory}

\indent

Consider the \textit{free} (neutral) scalar field $\vf$ of mass $m$
over $M_4$ subjected to the Klein--Gordon equation
\be
(\Box + m^2)\vf = 0,
\lbl{KG}
\ee
which is derivable from the action
\be
S_0(\vf) = \frac{1}{2} \int_{M_4} d^4x\,
 (\del_\nu\vf\,\del^\nu\vf - m^2 \vf^2)(x)
 =:  \int_{M_4} d^4x\, \L_0(x).
\lbl{cl}
\ee
To this classical field $\vf$ one may associate the quantum field
$\phi$, an operator valued distribution which acts on the bosonic Fock
space $\F$ built on the (spin~$0$, mass~$m$)-representation space of
the Poincar\'e group, under which the classical action \rf{cl} is
invariant. One has the commutation relations
\be
[\phi(x),\phi(y)] = i\,\Dl_{\mbox{\ssz JP}}(x-y)
 = i\Dl^+(x-y) - i\Dl^+(y-x),
\lbl{cr}
\ee
where the decisive locality (or microcausality) property
\be
[\phi(x),\phi(y)] = 0  \sepword{for}  x \sim y
\lbl{loc}
\ee
stems from the causal support property of the Jordan--Pauli
commutation ordinary distribution,
$\supp \Dl_{\mathrm{JP}} = \Bar V^+ \cup \Bar V^-$, and where
$(\Box + m^2) \Dl^+(x) = 0$. The problem is to make sense of the field
at the quantum level when $S_0$ is supplemented with an interaction
term
\be
S_{\mathrm{Int}}(\vf) = - \frac{\la}{4!} \int_{M_4} d^4x\, \vf^4(x)
 =: \int_{M_4} d^4x\, \L_1(x).
\lbl{Sint}
\ee
The first step consists in constructing a quantum version of $\vf^4$,
namely the fourth normal power $\wick{\phi^4}$ of~$\phi$, which will
share with the latter the causality property
\be
\left[\wick{\phi^4(x)},\wick{\phi^4(y)}\right] = 0
  \sepword{for} x \sim y.
\lbl{causality}
\ee
Let us recall that the necessity of defining Wick powers arises
because $\phi$ is a distribution and thus cannot be raised to a power.
The passage from $\phi^4$ to $\wick{\phi^4}$ is achieved by a suitable
renormalization
\be
\wick{\phi^4(x)}\ =\ \lim_{\row{x}{1}{4}\to x}
\wick{\phi(x_1)\cdots\phi(x_4)}
\lbl{fi4}
\ee
where the Wick-ordered product on the right hand side is given by the
usual Wick theorem for normal ordering products, \eg~\cite{BS80}. More
generally, one may associate to any local monomial
$\mono(\vf,\del\vf)(x)$ of the classical field $\vf$ and its
derivatives its Wick-ordered quantum version
$\wick{\mono(\phi,\del\phi)}(x)$. The subtraction rule is then more
complicated, but still relies on the construction of Wick-ordered
products~\cite{Sto71}.

\subsection{The general Epstein--Glaser construction}

\indent

As is well known, the definition of the interaction dynamics for the
quantum field leads to more severe renormalization problems.
Heuristically, one takes the Dyson formula for the scattering matrix,
\be
\Sb = T\,\exp\left( i\,\int_{M_4} d^4x\, \L_1(x) \right),
\lbl{Dys}
\ee
where $T$ denotes the chronological product or $T$-product. Due to the
distributional character of the free field $\phi$, equation~\rf{Dys}
is ill-defined. Our account of the idea of Stueckelberg--Bogoliubov--Shirkov
\cite{Stu,BS80}, further elaborated in~\cite{EG73}, to overcome this
problem goes as follows.

First, for $\L_1(x) = -\la\,\vf^p(x)/p!$, consider the extended
interacting Lagrangian
\be
\underline{g}(x) \underline{\L}(x)
 = \sum_{j=1}^p g_j(x)\,\L^{(j-1)}(x),
\lbl{Lint}
\ee
where $\underline{\L}$ denotes collectively a finite sequence of Wick
monomials of the free field and its derivatives, stable in the sense
that it contains all submonomials obtained by formal derivation of the
interacting Lagrangian $\L_1$ with respect to~$\phi$, so that in
particular the field $\L^{(p-1)} = \phi$ itself is there. The $g_j$
are Schwartz (smooth, rapidly decreasing) functions on~$M_4$ that
serve as sources to generate fields and composite fields, and also
play the role of Lorentz-covariant ``coupling constants''. They are
separated into $g_1$, which will be subject to the adiabatic limit,
namely $g_1 \to \la$, and the others, which will be sent to zero (note
that $g_p$ is the classical source ${\rm J}$ of $\phi$). In the
instance of $\phi^4_4$ theory, $\underline{\L}$ is given by the
following sequence of Wick powers of the free field $\phi$,
\be
\L^{(k)} = - \frac{\wick{\phi^{4-k}}}{(4-k)!}
  \sepword{for} k = 0,\dots,3, \qquad
\L^{(0)} = \frac{\L_1}{\la},  \quad  \L^{(4)}(x) = - \openone.
\lbl{L}
\ee

Second, one constructs a formal power series of the Gell-Mann--Low type in
the $g_j$ by defining the corresponding scattering operator $\Sb$
acting densely in $\F$,
\be
\Sb(\underline{g})
 =\ \openone + i \int_{M_4} d^4x\, \underline{g}(x) \underline{\L}(x)
 + \sum_{n\geq 2} \frac{i^n}{n!} \int_{M_4^n} d^4x_1\dots d^4x_n\
   T_n(x_1,\dots,x_n)\, \underline{g}(x_1)\cdots\underline{g}(x_n),
\lbl{S}
\ee
where the $T_n$ are operator-valued covariant symmetric distributions,
\be
T_n(\row{x}{1}{n})
 := T (\underline{\L}(x_1)\cdots \underline{\L}(x_n)).
\ee
These distributions will be properly defined time-ordered products of
the indicated Poincar\'e-invariant Wick monomials, whose recursive
construction is explained directly below. No convergence condition is
imposed on this formal expansion. For variants of the construction,
the reader is referred, as well as the original work by Epstein and
Glaser~\cite{EG73}, to the treatise~\cite{Scharf} and
to~\cite{Pinter}.

Third, following now the general strategy advocated by R. Stora
\cite{PR81,PoSt,Sto98}, we give a concise approach to the
Epstein--Glaser scheme in a geometrical setup which also applies to
the Euclidean manifold situation~\cite{unpub}.

Due to the symmetry of the distributional kernels in their arguments,
we can use the following shorthand.

\smallskip

\noindent
{\bf Notation}. For $n$ distinct points $x_i\in M_4$, $i = 1,\dots,n$,
one abbreviates 
\bea
(\row{x}{1}{n}) &=& X,\hcm n=|X|, \nn \\
\underline{g}(x_1)\cdots\underline{g}(x_n) &=& \underline{g}(X), \\
T_n(\row{x}{1}{n}) &=& T(X).  \nn
\eea

\begin{figure} [!ht]
\begin{center}
\parbox{42mm}{\begin{picture}(20,10)
\qbezier(0,0)(40,30)(80,0)
\qbezier(0,0)(-30,-30)(10,-20)
\qbezier(80,0)(110,-30)(70,-20)
\qbezier(10,-20)(40,-10)(70,-20)
\put(25,0){$\supp\underline{g}_1$}
\multiput(-30,-30)(10,0){15}{$-$}
\put(120,-30){$\mbox{Spacelike surface}$}
\put(40,-50){\oval(50,37)}
\put(25,-50){$\mbox{supp}\,\underline{g}_2$}
\put(17.5,-40){\line(-1,-2){30}}
\put(62.5,-40){\line(1,-2){30}}
\put(10,-90){$\supp\underline{g}_2 + \Bar{V}^-$}
\end{picture}}
\end{center}

\vskip 3cm

\caption{Causally separated domains}
\label{pict}
\end{figure}
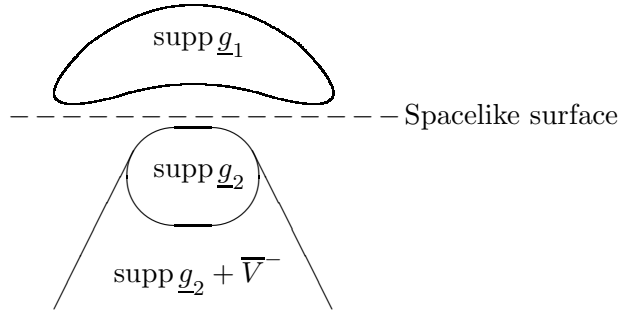

Formal scattering theory would yield
\be
T(X)\ = \sum_{\sg\in\P_{|X|}} \biggl(
 \prod_{k=1}^{|X|-1} \Th(x^0_{\sg(k)} - x^0_{\sg(k+1)}) \biggr)\,
 \underline{\L}(x_{\sg(1)}) \cdots \underline{\L}(x_{\sg(n)}),
\lbl{naive}
\ee
where $\P_{|X|}$ denotes the group of permutations of $|X|$ objects.
This na\"{\i}ve definition leads to the infinities of perturbation
theory because the operator-valued distributions in general cannot be
multiplied by the discontinuous step function. One requires instead
that if the coupling constant $\underline{g}$ splits into a sum of two
components with supports separated by a spacelike surface, (see
Figure~\ref{pict}),
then $\Sb(\underline{g})$ fulfils the \textit{causal factorization
property} \cite{BS80}, which stems from locality:
\be
\Sb(\underline{g}_1 + \underline{g}_2)
 = \Sb(\underline{g}_1)\,\Sb(\underline{g}_2)  \sepword{iff}
   \supp\underline{g}_1 \gtrsim \supp\underline{g}_2 \iff
   \supp\underline{g}_1 \cap (\supp\underline{g}_2 + \Bar V^-)
 = \emptyset.
\ee
This is woven into a double recursion hypothesis, as follows.

\indent

\noindent{\bf The double recursion hypothesis}.
Let $X$ be a finite subset of $|X|$ distinct points of $M_4$ and let
$M_4^{|X|}$ denote the cartesian product of $|X|$ copies of $M_4$.
As a $|X|$-uple, $X$ is a point in $M_4^{|X|}$.
Considering a partitioning 
$X = I \uplus I'$ of $X$ into two nonempty subsets $I$, $\emptyset
\subsetneqq I \subsetneqq X$ and~$I'=X\setminus I$ amounts to
projecting $X$ onto $I = 
(x_{i_1},\dots,x_{i_{|I|}})$, $(i_1,\dots,i_{|I|})\subsetneq
(1,\dots,|X|)$ with $1\leq |I| \leq |X|-1$. 
Let now $X$ vary in $M_4^{|X|}$ and look to the past from the region
defined by the complement $I'$, one gets a closed wedge in
$M_4^{|X|}$. Let $\Cc_I$ be the complement of this wedge, namely the
generalized open cone in $M_4^{|X|}$ defined ---see \rf{nopast}--- as
$$
\Cc_I = \set{X \in M_4^{|X|} : I \gtrsim I',\ I \uplus I' = X},
$$
which is translation-invariant. Then we make the following

\newcounter{hyp}
\begin{list}
{${}$\hs{4}\underline{Hypothesis} \arabic{hyp}}
{\usecounter{hyp}\labelwidth\setlength{\leftmargin}}
\item
(Causal factorization)
For $|X| < n$, $T(X)$ has been constructed so that
\be
T(X) = T(I)\,T(I') \sepword{in} \Cc_I.
\ee
\item
(Locality)
If $Y$ is another finite subset of $M_4$, then for $|X| < n$ and
$|Y| < n$, in the ${}$ \hs{13} region of $M_4^{\max(|X|,|Y|)}$ defined
by $X \sim Y = \set{X \gtrsim Y \mbox{ and } Y \gtrsim X}$,
\be
[T(X), T(Y)] = 0.
\ee
\end{list}
One sets $T(\emptyset) = 1$. For $n = 1$, since $\underline{\L}$ is
stable, the simplest such $T_1$ is $T_1 = \L^{(3)} = \phi$, the free
field itself, which obviously satisfies the recursion hypothesis
thanks to~\rf{loc}. It is to be remarked that although Hypothesis~1
implies $[T(I),T(I')] = 0$ in the region where $I \sim I'$,
Hypothesis~2 is actually necessary in order to obtain
$[\underline{\L}(x), \underline{\L}(y)] = 0$ for $x \sim y$, because
at $|X| = 1$ there is no causal factorization at all. Note, moreover,
that the definition of $\underline{\L}$ in \rf{Lint} is such that all
the $T$-products between all possible Wick monomials will be
recursively constructed.

The construction of $T(X)$ for $|X| = n$ is grounded on the following
lemma.

\begin{lem}[The geometrical lemma]
\label{geolem}
Given $|X|=n$ distinct points varying in $M_4$, with each partition $I
\uplus I' = X$ of 
$X$ into two non empty subsets, $I$ and~$I'$, one associates in
$M_4^{|X|}$ the translational invariant cone
$\Cc_I = \set{X \in M_4^{|X|} : I \gtrsim I',\ I \uplus I' = X}$. This
yields the open region
\be
\bigcup_{\emptyset\subsetneqq I\subsetneqq X} \Cc_I
 = M_4^{|X|} \setminus D_{|X|}\ ,
\ee
where $D_{|X|} = \set{x_1 = x_2 =\cdots= x_{|X|}}$ is the complete
diagonal in $M_4^{|X|}$.
\end{lem}

\noindent{\bf Proof}.
It goes by taking the complement, namely, showing that
\be
\bigcap_{\emptyset\subsetneqq I\subsetneqq X} \complement\,\Cc_I = D_{|X|}.
\lbl{hyp}
\ee
As a complement, $\Cc_I$ is written,
\be
\Cc_I = \set{X = I \uplus I',\  I \gtrsim I',\
  I \neq \emptyset,\ I' \neq \emptyset}
  = \bigcap_{i\in I,i'\in I'} \{x_i\gtrsim x_{i'}\}
  = \raisebox{-1ex}{\LARGE$\complement$}
    \bigcup_{i\in I,i'\in I'} \{x_i \leq x_{i'}\}.
\lbl{C}
\ee
The left hand side of \rf{hyp} is written as
\cbe
\bigcup_{\shortstack{\ssz $i(I)$\\ \ssz$i'(I')$}}
\bigcap_{\emptyset\subsetneqq I \subsetneqq X}
\{ i(I) \leq i'(I') \}
\cee
where the union runs over all the possible assignments
$I \mapsto i(I) \in I$ and $I \mapsto i'(I') \in I'$. Given such a
choice, say for $I'$, the intersection over all~$I$ is performed as
follows.
\begin{itemize}
\item
Step 1. Starting with $I = \{x_1\}$, for $I'$ one sets
$i'(I') = x_{i_2}$, this yields $x_1 \leq x_{i_2}$.
\item
Step 2. Taking $I = \{x_1,x_{i_2}\}$ and setting $i'(I') = x_{i_3}$
gives rise to
\cbe
\bigcup_{i(\{x_1,x_{i_2}\})}
 \blp \{x_1\leq x_{i_2}\} \cap \{i(\{x_1,x_{i_2}\})\leq x_{i_3}\} \brp
 = \{x_1\leq x_{i_2}\} \cap \{x_1\leq x_{i_3}\},
\cee
whatever the choice $i(\{x_1,x_{i_2}\})$ is, thanks to step~1.
\item
By iterating, at the $p$th step, $2 \leq p \leq |X|-1$, having
obtained ${\dps \bigcap_{k=2}^p \{x_1\leq x_{i_k}\}}$, one takes
$I = \{x_1,\dots,x_{i_p}\}$ and sets $i'(I') = x_{i_{p+1}}$, and
whatever is the choice $i(\{x_1,\dots,x_{i_p}\})$ one obtains
$$
\bigcup_{i(\{x_1,\dots,x_{i_p}\})} \blp\
 \bigcap_{k=2}^p \{x_1\leq x_{i_k}\} \cap \{i(\{x_1,\dots,x_{i_p}\})
 \leq x_{i_{p+1}}\}\brp = \bigcap_{k=2}^{p+1} \{x_1\leq x_{i_k}\}.
$$
After $|X|-1$ steps, one finally ends with
${\dps \bigcap_{k=2}^{|X|} \{x_1\leq x_k\}}$.
\item
One then chooses $I = \{x_2\}$, and defines $i'(I') = x_{j_2}$.
Proceeding as before, one gets after at most $|X|-1$ steps
${\dps \bigcap_{k=1,k\neq 2}^{|X|} \{x_2\leq x_k\}}$ containing the
restriction $x_2\leq x_1$. This yields
$$
\{x_1=x_2\} \cap \bigcap_{k=3}^{|X|} \{x_1\leq x_k\}.
$$
\item
Going further ahead, after a finite number of steps (at most
$\thalf(|X|-1)(|X|+2)$) one has exhausted the full diagonal
$D_{|X|} = \{x_1 =\cdots= x_{|X|}\}$. The latter remains stable under the
intersection with all the conditions $i(I)\leq i'(I')$ coming from all
unused~$I$.\qquad $\square$
\end{itemize}

The Lemma prompts us to define on $M_4^{|X|}$ the operator-valued
distributions
\be
T_I(X) = T(I)\,T(I'), \hcm I \uplus I' = X, \hcm
 I \neq \emptyset,\ I'\neq \emptyset.
\lbl{ext}
\ee
For $I$ and $J$ such that $\Cc_I \cap \Cc_J \neq \emptyset$, one
easily proves that on $M_4^{|X|}\,\setminus\,D_{|X|}$
\be
T_I(X)\Bigr|_{\Cc_I\cap\,\Cc_J} = T_J(X)\Bigr|_{\Cc_I\cap\,\Cc_J}.
\lbl{restrictIJ}
\ee
Indeed, using Hypothesis~1 in both $\Cc_I$ and $\Cc_J$,
$$
T(I)\,T(I') = T(I\cap J)\,T(I\cap J')\,T(I'\cap J)\,T(I'\cap J'),
$$
and similarly
$T(J)\,T(J') = T(J\cap I)\,T(J\cap I')\,T(J'\cap I)\,T(J'\cap I')$.
The equality of both expressions follows from
$[T(I\cap J'), T(I'\cap J)] = 0$ by applying Hypothesis~2 because in
$\Cc_I \cap \Cc_J$, $I \cap J' \sim I' \cap J$.

Due to the stability feature of $\underline{\L}$ the recursion
hypothesis is then reduced from operator-valued distributions to
scalar distributions according to the Wick expansion
formula~\cite{BS80} for $T$-products, which in our $\phi^4_4$ case is
typically written
\be
T \left(\frac{\wick{\phi^{p_1}(x_1)}}{p_1!}\cdots
        \frac{\wick{\phi^{p_n}(x_n)}}{p_n!}\right)
= \sum_{\shortstack{\ssz$q_j+r_j=p_j$\\ \ssz$j=1,\dots,n$}}
  \left< T \left( \frac{\wick{\phi^{r_1}(x_1)}}{r_1!} \cdots
   \frac{\wick{\phi^{r_n}(x_n)}}{r_n!} \right) \right>
  \frac{\wick{\phi^{q_1}(x_1)\cdots \phi^{q_n}(x_n)}}{q_1!\cdots q_n!}
\lbl{wick}
\ee
for $1 \leq p_j \leq 4$; we have followed standard practice of using
angle brackets to denote vacuum expectation values. In a shorthand
notation, in order to be more transparent in the accounting of all the
Wick monomials involved in $\underline{\L}$, this can be rewritten
using multi-indices $p,q,r$ with $|p| \leq 4|X|$ under the form
\be
T_p(X) = \sum_{q+r=p} \langle T_r(X)\rangle\,
 \frac{\wick{\phi^q(X)}}{q!}\ .
\lbl{wick'}
\ee
We are also using ``Theorem~0'' of \cite{EG73} which asserts that the
product of a translationally invariant scalar tempered distribution
given by Wick's theorem for vacuum expectation values with a Wick
product is a densely defined operator on the Fock space $\F$. For
instance, with $r_j = 1$ for all $j = 1,\dots,|X|$,
\be
\langle T_r(X) \rangle = \langle T\, \L^{(3)}(X) \rangle
 = \langle T\, \phi(x_1)\cdots\phi(x_{|X|}) \rangle
\ =\ \left\{ \begin{array}{cl}
       0, & \quad |X| \mbox{ odd}\\[3mm]
       {\dps \sum_{\shortstack{\ssz partitions\\ \ssz in pairs}}\
       \prod_{i<j} \langle T \phi(x_i)\phi(x_j) \rangle},
       & \quad |X| \mbox{ even}.
     \end{array} \right.
\lbl{Tpairing}
\ee

Therefore~\rf{restrictIJ} holds for the various scalar coefficients in
\rf{wick}. Then, on applying well-known results about the support of
distributions, the collection of restricted operator-valued
distributions
\be
\zer{T}_I^\0(X) = T_I(X)\bigr|_{\Cc_I}
\lbl{notI}
\ee
defines, thanks to Lemma~\ref{geolem}, a single operator-valued
distribution
\be
\zer{T}_p(X) = \sum_{q+r=p} \langle T_r(X) \rangle \,
\frac{\wick{\phi^q(X)}}{q!}
\lbl{not}
\ee
in $M_4^{|X|}\,\setminus\,D_{|X|}$. For instance, for $|X| = 2$ and
$|p| = 2$,
\be
\zer{T}_2^\0(x,y) = \wick{\phi(x)\phi(y)} + \Dl(x - y),
\lbl{T2}
\ee
where \rf{naive} has been used in order to define the Feynman
propagator $\Dl$,
\be
\langle T \phi(x)\phi(y) \rangle
 = i \Th(x^0 - y^0) \Dl^+(x - y) + i \Th(y^0 - x^0) \Dl^+(y - x)
 =: \Dl(x - y),
\lbl{propag}
\ee
with $i\Dl^+(x - y) = \langle \phi(x)\phi(y) \rangle$ being the
contraction and $(\Box + m^2) \Dl(x) = - i\,\dl(x)$.

\smallskip

Renormalization then consists in finding an extension $T(X)$ of
$\zer{T}(X)$ to the whole of $M_4^{|X|}$, namely extending from the
subspace of test functions whose support does not contain the diagonal
$D_{|X|}$ to the full space of test functions. Suppose that such an
extension $T(X)$ has been thus obtained for $|X| = n$. Then it is
clear that it satisfies the double recursion hypothesis: Hypothesis~1
by construction, by virtue of~\rf{ext}, and Hypothesis~2 because of
$T_I(X) - T_{I'}(X) = [T_I(X), T_{I'}(X)] = 0$ for $I \sim I'$ with
$I \uplus I' = X$, $I \neq \emptyset,\ I'\neq \emptyset$ by taking
$J = I'$ in \rf{restrictIJ}. Moreover, $T(X)$ for $|X| = n$,
constructed through the Wick's expansion~\rf{wick}, does fulfil Wick's
theorem.

\section{Extension of distributions in $\phi^4_4$ theory} \label{ext}

\indent

According to \rf{wick'}, the amplitudes $\langle T_r(X) \rangle$ are
scalar tempered-distribution coefficients in the Wick expansion in
terms of Wick monomials of the free field operators, belonging to
$\underline{\L}$. For a while, the former will be denoted by
$\zer{t\,}_r\!(X)$ so that \rf{not} is rewritten
\be
\zer{T}_p (X)\
 =\ \sum_{q+r=p} \zer{t\,}_r\!(X)\,\frac{\wick{\phi^q(X)}}{q!}\ .
\lbl{notp}
\ee
By virtue of Wick's theorem and \rf{propag}, $\zer{t\,}_r\!(X)$ is a
product of Feynman propagators at noncoinciding points and thus
translation-invariant; it depends on $|X| - 1$ independent difference
variables. According to Theorem~0 of~\cite{EG73}, extending
$\zer{T}(X)$ defined on $M_4^{|X|}\,\setminus\,D_{|X|}$ to the whole
$M_4^{|X|}$ amounts to extending each of the scalar distributions
$\zer{t\,}(X)$ to all of $M_4^{|X|}$. As discussed
in~\cite{Prange,Scharf,BrFr99,Mal}, an extension of a scalar
distribution $\zer{t\,}$ always exists, but it is of course not
unique, the ambiguity being given by a distribution with support on
the full diagonal $D_{|X|}$, whose degree depends on the ``degree of
singularity'' of $\zer{t\,}$. The power counting
theory~\cite{EG73,Scharf,BrFr99} then emerges in order to control
these ambiguities by shifting the singularity at the origin of
$\R^{4(|X|-1)}$, thanks to a change of variables from $X$ to
difference variables $(x_1 - x_2, \dots, x_{|X|-1} - x_{|X|})$.

\indent

\noindent
{\bf Power counting}.
The behaviour at the origin of $\R^{4n}$ of a scalar tempered
distribution $f$ can be described by both the scaling degree and the
singular order; the latter is also called the power counting index.

The scaling degree $\sg$ of a scalar distribution $f$ at the origin of
$\R^{4n}$ is defined to be
\be
f\in \D'(\R^{4n}),\qquad
\sg(f) = \inf \set{s : \lim_{\la\to 0} \la^s f(\la x) = 0},
\ee
where the limit is taken in the sense of distributions. It is clearly
a generalization of the notion of degree of a homogenous distribution.
The concept of scaling degree is due to Steinmann~\cite{OS}; its main features
are summed up in the following statement of~\cite{BrFr99}.

\begin{lem}
\label{scaling}
For $f \in \D' (\R^{4n})$ and $a$ a multi-quadri-index,
\begin{enumerate}
\item $\sg(x^a f) = \sg(f) - |a|$.
\item $\sg(\del_a f) = \sg(f) + |a|$,
      (derivation increases the scaling degree).
\item If $\aux \in \D(\R^{4n})$ with $\sg(\aux) \leq 0$, then
      $\sg(\aux\,f) \leq \sg(f)$.
\item $\sg(f_1\ox f_2) = \sg(f_1) + \sg(f_2)$.
\end{enumerate}
\end{lem}

The singular order $\om$ of the scalar distribution $f$ at the origin
of $R^{4n}$ is defined to be
\be
f \in \D' (\R^{4n}),\qquad  \om(f) = [\sg(f) - 4n],
\lbl{om}
\ee
where $4$ is the spacetime dimension and $[\cdot]$ denotes integer
part.

In particular, the amplitudes will be scalar tempered distributions
and, for instance, the propagator
$\langle T\phi(x)\phi(y)\rangle = \Dl(x-y)$ in the $\phi^4_4$ theory
has scaling degree $\sg(\Dl) = +2$, and since the scaling degree of
the tensor product of scalar distributions is the sum of those of the
factors, in particular the scaling degree of
${\dps \prod_{k=1}^{|X|-1} \Dl(x_k - x_{k+1})}$ is $2(|X| - 1)$.

In $\R^4$, for the Dirac distribution $\sg(\dl) = 4$, thus
$\om(\dl) = 0$, and for the following tempered distributions,
\be
\om(\Dl^+) = \om(\Dl) = - 2, \qquad
\om\bigl( \Dl(x_1 - x_2)^{|X|-1} \bigr) = 2|X| - 6.
\lbl{index}
\ee
Moreover, if $\om(\zer{t\,}) < 0$, then there is a unique scalar
distribution $t \in \D'(\R^{4n})$ which extends $\zer{t\,}$ from the
subspace $\D(\R^{4n}\setminus\{0\})$ of test functions whose support
does not contain the origin, in the sense that
$\<t,\vf> = \<\zer{t\,},\vf>$ for any
$\vf \in \D(\R^{4n}\setminus\{0\})$.

For $\om := \om(\zer{t\,})\geq 0$, call $\D_{\om}(\R^{4n})$ the
finite-codimensional subspace of test functions vanishing up to order
$\om$ at the origin,
\be
\D_{\om}(\R^{4n})
 := \set{\vf \in \D(\R^{4n}) : \del_{a}\vf(0) = 0,\
         \forall\ |a|\leq \om},
\ee
where $a$ is a multi-quadri-index.

For all $\zer{t\,}\in\D'_{\om}(\R^{4n})$, there exist scalar
distributions $t \in \D'(\R^{4n})$ with $\om(t) = \om$ and
$\<t,\vf> = \<\zer{t\,},\vf>$ for any $\vf \in \D_{\om}(\R^{4n})$.
Power counting theory asserts that there exists a \textit{minimal
class} of extensions, (in the sense that their singular orders are the
smallest possible), within which the ambiguity on the extensions
reduces to the form
\be
\Dl t(x)
 = \sum_{|a|=0}^\om \frac{(-1)^{|a|}\,C^{a}}{a!}\ \dl^{(a)}(x),
\lbl{ambig}
\ee
where $\<(-1)^{|a|}\,\dl^{(a)},\vf> = \del_a \vf (0)$ and where the
$C^{a}$ are free constants due to translation invariance, 
which are often fixed by symmetry (Lorentz, gauge\dots) considerations,
with the help of Ward and/or Slavnov identities,
\eg~\cite{Scharf,Pinter,Dut,Prangebis}. Note that
$\om(\Dl t) = \om(t)$. In particular in $\phi^4_4$ theory, assume that
a scalar extension has been constructed for each of the $\zer{t\,}_r$
in \rf{notp} with singular degree $\om(r)\geq 0$; then the ambiguities
\rf{ambig} yield an ambiguity for the extension of~\rf{notp},
\be
\Dl T_p(X) = \sum_{q+r=p}
 P_r(\del)\, \dl(x_1 - x_2) \cdots \dl(x_{|X|-1} - x_{|X|})
\, \frac{\wick{\phi^q(X)}}{q!}
\ee
where each of the polynomials of derivatives (with constant
coefficients)
\be
P_r(\del) \ =\ \sum_{|a|=0}^{\deg P_r}
\frac{(-1)^{|a|}\, C^{a}}{a!}\ \del_{a}
\ee
has (maximal) degree $\om(r)$ given by power counting according to
\be
\deg P_r \leq \om(r) = \om_X - \om\Bigl(\frac{\wick{\phi^q}}{q!}\Bigr),
\hcm
\om_X = \sum_{j=i}^{|X|}
\left( \om\Bigr(\frac{\wick{\phi^{p_j}}}{p_j!}\Bigl) - 4 \right) + 4.
\lbl{div}
\ee
The power counting index (or degree) $\om$ for Wick monomials is
defined to be
\be
\om\Bigr(\frac{\wick{\phi^q}}{q!}\Bigl) = \om(\phi)\,|q|.
\ee
For the case of interest of a massive scalar free field,
$\om(\phi) = 1$, coming from
\be
\om(\phi)\ =\ \thalf(4 - \deg K_0),
\ee
a formula which stems from Lemma~\ref{scaling} together with the
requirement that the singular degree of the free Lagrangian
$\L_0 = \wick{\phi\,K_0\phi}$ be~4.

Notice now that $\langle T_r(X) \rangle$ is represented as a sum of
Feynman graphs $\Ga$ (connected or not) with $|X|$ vertices and
$|q| = |p| - |r|$ external lines, where $q_j$ external lines are
attached to the $j$th vertex of type $\L^{(4-p_j)}$,
$j = 1,\dots,|X|$, so that
$\om(r) = {\dps \sum_{j=i}^{|X|}
\left(\om\Bigl(\frac{\wick{\phi^{r_j}}}{r_j!}\Bigr)-4\right) + 4}$ is
nothing but the \textit{superficial degree of divergence} $\om(\Ga)$
of the corresponding graphs $\Ga$ giving the UV behaviour,
\eg~\cite{IZ}.

\smallskip

For instance, according to \rf{wick'}, if $|p| = 4|X| = |r| + |q|$,
i.e., all the $|X|$ vertices are of the type
$\L^{(0)} = -\wick{\phi^4}/4!$, owing to \rf{div} one has
$\om(\Ga) = 4 - E$, where $E = |q|$ is the number of external lines
associated to the bosonic field $\phi$. So in $\phi^4_4$ theory, only
graphs with 2 or 4 external legs will be superficially divergent.
(Vacuum graphs $(E = 0)$ are dropped because they have been shown to
vanish in the adiabatic limit~\cite{Scharf,Dut}.) Note that in their
definition, the amplitudes $\langle T_r(X) \rangle$ require the
knowledge of the various $T$-products $T_r(X)$ between all the Wick
monomials belonging to $\underline{\L}$.

\smallskip

We shall now concentrate in full generality on a given graph $\Ga$
with $|X|$ vertices which contributes to the amplitude
$\langle T_r(X) \rangle$ (a tempered scalar distribution) and set
$f(\Ga) := \zer{t\,}_r\!(X)$. It remains to recall how an extension of
$f(\Ga)$ can be exhibited by a Taylor-like subtraction, anticipated in
the heuristic discussion in~Section~\ref{motiv}, the so-called
$W$-operation \cite{EG73,Scharf}.

According to both a given graph $\Ga$ containing $n=|X|$ vertices and
having power counting index $\om:=\om(\Ga)$, and a choice of a
function $\aux_\Ga$ of  $n - 1$ variables such that $\aux_\Ga(0) = 1$
and $\del_a \aux_\Ga(0) = 0$ with $a$ a multi-quadri-index such that
$1 \leq |a| = |a_1| + \cdots + |a_{n-1}| \leq \om$, one defines the
mapping
$$
W_\Ga \ :\ \D(M_4^n) \lra \D_{\om}(M_4^n), \qquad
  \vf \longmapsto W_\Ga\,\vf $$
where
\begin{equation}
\left(W_\Ga\,\vf\right)(X)
 = \vf(X) - \aux_\Ga(x_1-x_2,\dots,x_{n-1}-x_n) \sum_{|a|=0}^\om
   \frac{(x_1-x_2)^{a_1}\cdots(x_{n-1}-x_n)^{a_{n-1}} }{a!}\ 
\del_a \vf \bigr|_{D_{|X|}}.
\lbl{W}
\end{equation}

\begin{lem}
\label{proj}
For each superficially divergent graph $\Ga$, $\om(\Ga) \geq 0$,
$W_\Ga$ is a projector, $W_\Ga (1 - W_\Ga) = 0$.
\end{lem}

A subtraction operation can be defined by $S_\Ga := 1 - W_\Ga$.

In the Epstein--Glaser scheme, to any choice $I \subseteq X$ of
vertices of $\Ga$ there corresponds a subgraph $\ga$ of~$\Ga$.

\begin{lem}
For any subgraph $\ga \subseteq \Ga$ with $|I|$ vertices,
$|I| \leq |X|$ and $\om(\ga) \geq \om(\Ga)$, one has
$S_\Ga W_\ga = 0.$
\end{lem}

By transposition one defines the action of these projectors on
tempered distributions. We keep then the same names for them; but
beware that the order of operators in formulae like those of the lemma
is inverted.

Of course the extension may be fixed if desired; namely, for a given
set of the $C$~constants and a given $\aux$ in \rf{ambig}, there is a
unique extension $\tilde f(\Ga)$ such that
\begin{eqnarray*}
&{}& \<\tilde f(\Ga),\vf>
 = \<f(\Ga),W_\Ga\vf>, \qquad \forall\ \vf\in \D(M_4^n) \\[2mm]
&{}& \<\tilde f(\Ga),(x_1-x_2)^{a_1}\cdots(x_{n-1}-x_n)^{a_{n-1}}
\aux_\Ga(x_1-x_2,\dots,x_{n-1}-x_n) >
 = C^a, \quad 0\leq |a| \leq \om(\Ga).
\end{eqnarray*}

\smallskip

Thanks to the causal factorization property involved in the
definitions \rf{ext} and \rf{notI}, the chronological products
$T(I)\,T(I')\bigr|_{\Cc_I}$ have already been extended to subdiagonals
of $M_4^{|X|}$ by construction. We call the extension to these
subdiagonals the \textit{renormalization of subdivergences} and the
extension to the complete diagonal in the last step the
\textit{renormalization of the overall divergence}. The known
subtraction-transposition procedures provide one way to solve this
extension problem. All terms in $\phi^4_4$ theory of $\Sb$ up to
order~3 in the vertices can be found in~\cite{Pinter}, to which the
reader is referred. We also remit to~\cite{Pinter} for the bit of
microlocal theory required to cross-check that the chronological
products are well defined. It is emphasized that in the CPT scheme grounded on
locality the arduous problem of overlapping divergences is
disentangled in a recursive way within an 
operator formalism before the adiabatic limit is taken.

\noindent{\bf A criterion} \cite{EG73,Sto98}.
The vacuum
expectation values of operator-valued distributions are conveniently
represented by Feynman graphs. In order to recognize that a 
finite expression for a Feynman diagram is actually a renormalization,
one proceeds in configuration space by checking that in \textit{all}
the open cones $\Cc_I$ associated to 
the splittings $X = I\uplus I'$, $I \neq \emptyset$,
$I'\neq \emptyset$, the amplitude factorizes as
\be
\langle T(X)\rangle\Bigr|_{\Cc_I} \
 =\ \langle T_I(X)\rangle\ \langle T_{I'}(X)\rangle
  \prod_{\shortstack{\ssz$i\in I$,\ssz$i'\in I'$\\ \ssz$(ii')\in L$}}
\langle \phi(x_i)\phi(x_{i'})\rangle .
\lbl{crit}
\ee
Here $\langle T_I(X)\rangle$ and $\langle T_{I'}(X)\rangle$ are
renormalized amplitudes (subgraphs of respective orders $|I|$
and~$|I'|$) computed at lower orders, and the product ranges over the
internal lines $(ii')$ in the set $L$ of internal lines connecting the
vertices $i\in I$ and $i'\in I'$. Formula \rf{crit} directly follows
by applying Wick's theorem to the product in~\rf{not}.

\smallskip

At this stage, an important remark (chiefly due to D.~Kreimer and R. Stora)
needs to be made: Hopf algebras are
naturally related to problems in combinatorics and shuffle theory.
There is thus a Hopf algebra structure \textit{already} implicit in
the recursive method of Epstein and Glaser, in the geometric
presentation advocated here. That algebra describes the stratification
of the short-distance singularities along
subdiagonals~\cite{DirkPoly}. However, to keep close to the
treatment in~\cite{CoKrI} ---and to the standard language--- we
formulate our Hopf algebra in terms of Feynman diagrams. This has the
drawback that the algebraic method appears, although perhaps elegantly,
almost as an
afterthought; but we gladly pay this price in order to have a more
pictorial description. For practical purposes, then, Feynman graphs
serve as mnemonic devices to keep the accounting of the amplitudes and
their singularities straight (for memory: vertices correspond to
coupling constants, external lines to field insertions and internal
lines to vacuum expectation values). Also in this connection, we should
mention that there is also a Hopf algebra structure~\cite{Fauser} behind 
Wick ordering, on which we lightly treaded in Section~3. We intend to
return to these matters in a sequel article.

\section{The counterterm map in Epstein--Glaser renormalization}

\indent

We conveniently begin by collecting all the ``graphological''
definitions needed in this section and the next. As already mentioned,
a \textit{graph} or diagram of the theory is specified by a set of
\textit{vertices} and a set of \textit{lines} among them;
\textit{external} lines are attached to only one vertex each,
\textit{internal} lines to two. We mentioned, once more, that diagrams with
no external lines need not be taken into account. In $\phi^4_4$
theory, only graphs with an even number of external lines are to be
found; apart from the interaction vertices with joining of four lines,
we shall consider two-line self-energy vertices corresponding to mass
insertions.

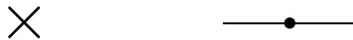
\begin{figure}[!hbp]

\indent

\begin{center}
$
\mbox{\Huge $\times$}
\qquad\qquad\qquad
\parbox{15mm}{\begin{picture}(20,10)
\put(0,8){\line(1,0){50}}
\put(22.5,5.5){$\bullet$}
\end{picture}}
$
\end{center}
\caption{$\wick{\phi(x)^4}$ interaction vertex and a two-line
self-energy vertex.}
\label{fig1}
\end{figure}

\noindent
A diagram is connected when any two of its vertices are joined by
lines, of course.

Given a graph $\Ga$, a \textit{subdiagram} $\ga$ of $\Ga$ is specified
by a subset of at least two vertices of $\Ga$ and a subset of the
lines that join these vertices in $\Ga$. Clearly, the external lines
for $\ga$ include not only the subset of original incident lines but
some internal lines of $\Ga$ not included in $\ga$. By exception, the
empty subset $\emptyset$ will be admitted as a subdiagram of $\Ga$. As
well as $\Ga$ itself.

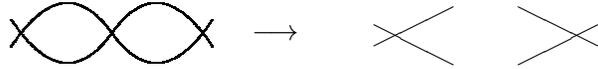
\begin{figure}[!hbp]

\indent

\begin{center}
$\ppois \qquad \lra \qquad
\parbox{15mm}{\begin{picture}(20,10)
\put(0,0){\line(2,1){30}}
\put(0,8){\line(2,-1){30}}
\end{picture}}
\qquad
\parbox{15mm}{\begin{picture}(20,10)
\put(20,0){\line(-2,1){30}}
\put(20,8){\line(-2,-1){30}}
\end{picture}}
$
\end{center}

\indent

\caption{\small Example of a one-vertex reducible (though proper)
graph split into a disconnected graph}
\label{fig2}
\end{figure}

The connected pieces of $\Ga$ are the maximal connected subdiagrams. A
diagram is \textit{proper}, i.e., 1PI, when the number of its
connected pieces would not increase on the removal of a single
internal line; otherwise it is called \textit{improper}, i.e., 1PR. An
improper graph is the union of its proper components plus subdiagrams
containing a single line. Note that in the theory at hand there are
proper diagrams, like the ``bikini''
of Figure~\ref{fig2}, which are however one-\textit{vertex} reducible,
i.e., they are made more disconnected on the removal of a vertex.

A \textit{subgraph} of a proper graph is a subdiagram that contains
all the propagators that join its vertices in the whole graph; as
such, it is determined solely by the vertices. A subgraph of an
improper graph $\Ga$, distinct from $\Ga$ itself, is a proper
subdiagram each of whose components is a subgraph with respect to the
proper components of $\Ga$: in other words, a product of subgraphs. We
write $\ga \subseteq \Ga$ if and only if $\ga$ is a subgraph of $\Ga$
as defined: this is the really important concept for us.

Two subdiagrams $\ga_1,\ga_2$ of $\Ga$ are said to be
\textit{nonoverlapping} when $\ga_1 \cap \ga_2 = \emptyset$ or
$\ga_1 \subset \ga_2$ or $\ga_2 \subset \ga_1$ in an obvious sense;
otherwise they are overlapping. Given $\ga \subseteq \Ga$, the
quotient graph or cograph $\Ga/\ga$ is defined by shrinking $\ga$ in
$\Ga$ to a vertex, that is to say, $\ga$ (bereft of its external
lines) is considered as a vertex of $\Ga$, and all the lines in $\Ga$
not belonging to (amputated) $\ga$ belong to $\Ga/\ga$.

Bogoliubov's $R$-operation is defined in the context of CPT by the
$W$-operation of the previous section. To be precise, let
$f(\Ga) \in \D'(\R^{4k}\setminus D)$, with $k$ the number of vertices
and $D$ any ``dangerous'' diagonals, the scalar distribution
corresponding, by the na\"{\i}ve rule for the time-ordered product of
propagators, to a diagram $\Ga$ contributing to the scattering matrix
at order $k$, with $E$ bosonic field insertions associated to the
external lines. We can and shall assume here that $\Ga$ is connected,
since clearly
\be
f(\ga_1 \uplus\cdots\uplus \ga_n) = f(\ga_1) \dots f(\ga_n).
\ee
We know that $f$ actually depends only on $k-1$ difference variables.
If $\Ga$ is superficially divergent (i.e., if $\om_\Ga:=\om(f(\Ga))\geq0$)
and has subdivergences, we let
\be
R_\Ga f(\Ga) = W_\Ga\, \bar R_\Ga f(\Ga),
\ee
where the operation $\bar R_\Ga$ reflects the renormalization of the
subdivergences present in $f(\Ga)$. More precisely, denoting by
$\vf_\Ga$ the product of the Wick monomial and the coupling constants
corresponding to $\Ga$, and keeping in mind the convention that the
same symbol denotes a linear map of the theory and its transpose,
\be
\<R_\Ga f(\Ga), \vf_\Ga> = \<\bar R_\Ga f(\Ga), W_\Ga\vf_\Ga>
 = \<\bar R_\Ga f(\Ga), (1-S_\Ga)\vf_\Ga>,
\lbl{renorm}
\ee
which, with the definition of the overall ``counterterm''
$C(\Ga) := - S_\Ga\, \bar R_\Ga f(\Ga)$, is rewritten as
\be
\<R_\Ga f(\Ga), \vf_\Ga> = \<\bar R_\Ga f(\Ga) + C(\Ga), \vf_\Ga>.
\ee

Now, Bogoliubov's recursive formula for $\bar R_\Ga$ is known to be
\begin{eqnarray}
\bar R_\Ga f(\Ga)
 := f(\Ga) - \sum_{\emptyset\subsetneqq\ga\subsetneqq\Ga}
    \left( S_\ga\, \bar R_\ga f(\ga) \right) f(\Ga/\ga),
\lbl{bog}
\end{eqnarray}
where the sum is taken over all \textit{proper, superficially
divergent, not necessarily connected subgraphs}, and $f(\Ga/\ga)$ is
just defined by the splitting $f(\ga)f(\Ga/\ga) = f(\Ga)$. Then
$\bar R_\ga$ and also $C(\ga) := -S_\ga\bar R_\ga f(\ga)$ are
recursively defined. It is clear that the Epstein--Glaser method, as
described in the two previous sections fits perfectly with
Bogoliubov's recursion; in fact, it was tailor-made for the purpose.

A graph without subdiagrams of the kind just defined is called
\textit{primitive}. At this point, a clarification is perhaps needed:
the previous definition automatically banishes subdiagrams that result
in codiagrams with ``loop lines'', whose initial and final vertices
coincide: the so-called tadpoles. This is natural in our context,
since tadpoles do not appear in the (unrenormalized) $\Sb$-matrix
expansion in real space. In particular, \textit{the setting sun
diagram is primitive} here, as it does not possesses proper
\textit{subgraphs}. Had we allowed general subdiagrams in place of
subgraphs, that would not be the case. However, as explained by Hepp
in~\cite[pp.~481--483]{Hepp71}, the added terms are irrelevant. A
(nontrivial) proof that the inclusion, either uniquely of subgraphs or
of general subdiagrams, in the $\bar R$ operation leads to equivalent
results is given in the context of the parametric representation
in~\cite{BZuber}. (``Tadpoles with tail'' corresponding to subdiagrams
with a single external ---for them--- incident line do not occur at all in
$\phi^4_4$ theory.)

Henceforth, we shall say simply ``divergent'' to mean ``superficially
divergent''.

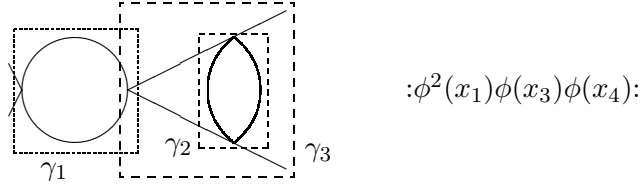
\begin{figure}[!ht]

\vskip 1cm

\begin{center}
\parbox{42mm}{\begin{picture}(20,10)
\put(-3,-19){\dashbox{1}(47,47)}
\put(7,-27){$\ga_1$}
\put(0,5){\line(-1,2){5}}
\put(0,5){\line(-1,-2){5}}
\put(20,5){\circle{40}}
\put(40,5){\line(2,1){60}}
\put(40,5){\line(2,-1){60}}
\put(37,-28){\dashbox{3}(66,66)}
\put(107,-20){$\ga_3$}
\put(67,-17){\dashbox{2}(26,43)}
\put(54,-18){$\ga_2$}
\qbezier(70,5)(70,17)(80,25)
\qbezier(70,5)(70,-7)(80,-15)
\qbezier(90,5)(90,17)(80,25)
\qbezier(90,5)(90,-7)(80,-15)
\end{picture}}
$\qquad \wick{\phi^2(x_1)\phi(x_3)\phi(x_4)}$
\end{center}

\vskip .7cm

\caption{Identification of proper subdiagrams.}
\label{fig4}
\end{figure}

Before tackling the main result, let us work out in full detail an
example, to see how the recursive definition works in practice in the
CPT framework. Suppose that the chronological product \rf{wick'} has
been constructed at the fourth order, and let us consider the densily
defined operator $\<{T}(X),\underline{g}(X)>$.  Among all the possible
terms in the Wick expansion we pick first the one corresponding to the
graph of Figure~\ref{fig4}. Let us choose and fix once and for all the
auxiliary function $\aux_\ga$ for each type of superficially divergent
subdiagram $\ga$; we shall always take $\aux$ even. There is a mass
scale associated naturally to each of them, which we can suppose
``universal'' in the style of 't Hooft and Veltman, if convenient. The
singular order or subtraction degree $\om(\ga)$ is here zero in all
cases. In general, if $\ga = \ga_1 \uplus\cdots\uplus \ga_n$ is a
product of connected pieces, we take $\aux_\ga =
\aux_{\ga_1}\cdots\aux_{\ga_n}$. The (sub)space on which the
subtraction takes place will be always explicitly displayed using
coordinates. Numbering the vertices from left to right, one gets the
following operator contributing to the fourth order,
\bea 
\la^4 \
\<\bigl[\Dl^2(x_1-x_2)\Dl(x_2-x_3)\Dl(x_2-x_4)\Dl^2(x_3-x_4)\bigr]_R,
\varphi_\Ga(x_1,x_2,x_3,x_4)>, \eea where
$\varphi_\Ga(x_1,x_2,x_3,x_4)$ stands for
$\wick{\phi^2(x_1)\phi(x_3)\phi(x_4)}\ g(x_1)g(x_2)g(x_3)g(x_4)$.
There are four subdiagrams to consider: the fish on the left, called
$\ga_1$; the fish on the right, $\ga_2$; the whole ice-cream cone on
the right, $\ga_3$; and the union of the two fish, $\ga_1 \uplus
\ga_2$. Note that $\ga_1$ and $\ga_3$ overlap, $\ga_1 \cap \ga_3 =
\{x_2\}$. See again Figure~\ref{fig4} for these subdiagrams.

The cardinality of the maximal chains $\ga_2 \subset \ga_3 \subset\Ga$
or $\ga_1,\ga_2 \subset \ga_1 \uplus \ga_2 \subset \Ga$, corresponding
to the maximal number of nested subdiagrams, counting $\Ga$, is~3.
This number we can call the \textit{depth} of the diagram (the
stratification of short-distance singularities is ordered by depth).

One of the things that make life with CPT truly comfortable is that,
by the simple magic of duality, one actually proceeds from the top
down, taking care first of the overall divergence, and then of the
subdivergences as they unfold~\cite{Pinter}. Working in coordinate
space, moreover, makes rather intuitive the distribution-theoretic
reasoning. There is no doubt about the soundness of the operations
performed, since we work with very well-behaved Schwartz functions.
With a notation that we hope is self-evident, we compute with the help
of \rf{W}
\bea
\leqa{ \<\bigl[\Dl^2(x_1-x_2)\Dl(x_2-x_3)\Dl(x_2-x_4)
 \Dl^2(x_3-x_4)\bigr]_R,
\wick{\phi^2(x_1)\phi(x_3)\phi(x_4)}g(x_1)g(x_2)g(x_3)g(x_4)>}
\nn\\
&& = \langle \bigl[ \Dl^2(x_1-x_2)\Dl(x_2-x_3)\Dl(x_2-x_4)
 \Dl^2(x_3-x_4) \bigr]_{\bar R},
\wick{\phi^2(x_1)\phi(x_3)\phi(x_4)}g(x_1)g(x_2)g(x_3)g(x_4)
\nn\\
&&\hs{60} -\ \aux_{\Ga}(x_1-x_2,x_2-x_3,x_3-x_4)
  \wick{\phi^4(x_4)}g^4(x_4) \rangle.
\nn
\eea
According to \rf{bog}, this is now reexpressed as
\bea
\leqa{\langle \Dl^2(x_1-x_2)\Dl(x_2-x_3)\Dl(x_2-x_4)\Dl^2(x_3-x_4),
\wick{\phi^2(x_1)\phi(x_3)\phi(x_4)}g(x_1)g(x_2)g(x_3)g(x_4)}
\nn \\
&& - \aux_{\Ga}(x_1-x_2,x_2-x_3,x_3-x_4)\wick{\phi^4(x_4)}g^4(x_4)
- \aux_{\ga_1}(x_1-x_2)\wick{\phi^2(x_2)\phi(x_3)\phi(x_4)}
   g^2(x_2)g(x_3)g(x_4)
\nn \\
&&\hs{30} + \aux_{\ga_1}(x_1-x_2)\aux_{\Ga}(0,x_2-x_3,x_3-x_4)
 \wick{\phi^4(x_4)}g^4(x_4) \rangle
\nn \\
&&- \langle\Dl^2(x_1-x_2)\Dl^2(x_2-x_4)\Dl^2(x_3-x_4),
\aux_{\ga_2}(x_3-x_4)\wick{\phi^2(x_1)\phi^2(x_4)}g(x_1)g(x_2)g^2(x_4)
\nn \\
&&\hs{30} - \aux_{\ga_2}(x_3-x_4)\aux_{\Ga}(x_1-x_2,x_2-x_3,0)
\wick{\phi^4(x_4)}g^4(x_4) \rangle
\nn \\
&& - \langle\Dl^2(x_1-x_2)[\Dl(x_2-x_3)\Dl(x_2-x_4)
      \Dl^2(x_3-x_4)]_{\bar R},
\nn\\
&&\hs{30}
\aux_{\ga_3}(x_2-x_3,x_3-x_4)\wick{\phi^2(x_1)\phi^2(x_4)}
  g(x_1)g^3(x_4)
\nn \\
&&\hs{30} - \aux_{\ga_3}(x_2-x_3,x_3-x_4)\aux_\Ga(x_1-x_2,0,0)
\wick{\phi^4(x_4)}g^4(x_4)\rangle
\nn \\
&&- \langle [\Dl^2(x_1-x_2)\Dl^2(x_3-x_4)]_{\bar R}\Dl^2(x_2-x_4),
\nn\\
&&\hs{30} \aux_{\ga_1\uplus\ga_2}(x_1-x_2,x_3-x_4)
\wick{\phi^2(x_2)\phi^2(x_4)}g^2(x_2)g^2(x_4)
\lbl{9} \\
&&\hs{30} - \aux_{\ga_1\uplus\ga_2}(x_1-x_2,x_3-x_4)
 \aux_\Ga(0,x_2-x_3,0) \wick{\phi^4(x_4)}g^4(x_4)\rangle.
\lbl{10}
\eea
That corresponds to rewriting
\be
\<\bar R_\Ga f(\Ga),W_\Ga\vf_\Ga>
 = \biggl<f(\Ga) - \sum_{\emptyset\subsetneqq\ga\subsetneqq\Ga}
    \left(S_\ga\, \bar R_\ga f(\ga)\right) f(\Ga/\ga),
     (1-S_\Ga)\vf_\Ga \biggr>
\ee
as
\be
\sum_{\emptyset\subseteqq\ga\subsetneqq\Ga}
\<\bar R_\ga f(\ga), -S_\ga f(\Ga/\ga)(1-S_\Ga)\vf_\Ga>,
\ee
where we agree that $-S_\emptyset = 1$. Note that
$\bar R_{\ga_1}f(\ga_1) = f(\ga_1)$,
$\bar R_{\ga_2}f(\ga_2) = f(\ga_2)$, because these subgraphs have no
subdivergences. At this stage, however, we are not yet finished, since
there are still the subgraphs $\ga_2 \subset \ga_3$ and
$\ga_1,\ga_2 \subset \ga_1 \uplus \ga_2$ to be taken into account.
When this is done, owing to \rf{bog} we obtain, in addition to the ten
terms already displayed (with the last square brackets on the left
hand side definitively gone), six more terms of the form
\bea
&&\hs{-20} \langle \Dl^2(x_1-x_2)\Dl^2(x_2-x_4)\Dl^2(x_3-x_4),
\aux_{\ga_2}(x_3-x_4)\aux_{\ga_3}(x_2-x_3,0)
\wick{\phi^2(x_1)\phi^2(x_4)}g(x_1)g^3(x_4)
\nn \\
&-& \aux_{\ga_2}(x_3-x_4)\aux_{\ga_3}(x_2-x_3,0)
 \aux_{\Ga}(x_1-x_2,0,0) \wick{\phi^4(x_4)}g^4(x_4)
\nn\\
&+& \aux_{\ga_1}(x_1-x_2)\aux_{\ga_1\uplus\ga_2}(0,x_3-x_4)
\wick{\phi^2(x_2)\phi^2(x_4)}g^2(x_2)g^2(x_4)
\lbl{3} \\
&-&
\aux_{\ga_1}(x_1-x_2)\aux_{\ga_1\uplus\ga_2}(0,x_3-x_4)
\aux_{\Ga}(0,x_2-x_3,0) \wick{\phi^4(x_4)}g^4(x_4)
\lbl{4} \\
&+& \aux_{\ga_2}(x_3-x_4)\aux_{\ga_1\uplus\ga_2}(x_2-x_3,0)
\wick{\phi^2(x_2)\phi^2(x_4)}g^2(x_2)g^2(x_4)
\lbl{5} \\
&-& \aux_{\ga_2}(x_3-x_4)\aux_{\ga_1\uplus\ga_2}(x_1-x_2,0)
\aux_{\Ga}(0,x_2-x_3,0)\wick{\phi^4(x_4)}g^4(x_4)\rangle.
\lbl{6}
\eea
But note now that the terms \rf{3} and \rf{5} are
actually equal and, in view of the assumed factorization property for
$\aux_{\ga_1\uplus\ga_2}$, either of them cancels the term
\rf{9} above; and analogously for terms
\rf{4} and \rf{6} and \rf{10}, respectively.
Symbolically, $(-S_{\ga_1\uplus\ga_2})(-S_{\ga_1})
 = (-S_{\ga_1\uplus\ga_2})(-S_{\ga_2}) = S_{\ga_1\uplus\ga_2}
 = S_{\ga_1}S_{\ga_2}$, so the pattern of the subtraction, that was
\be
f(\Ga) \mapsto (1 - S_{\Ga})
[1 - S_{\ga_1} - S_{\ga_2} - S_{\ga_3}(1 - S_{\ga_2})
   -  S_{\ga_1\uplus\ga_2} (1 - S_{\ga_1} - S_{\ga_2})] \,f(\Ga),
\ee
becomes
\be
f(\Ga) \mapsto (1 - S_{\Ga})
(1 - S_{\ga_1} - S_{\ga_2} - S_{\ga_3}(1 - S_{\ga_2})
 + S_{\ga_1}S_{\ga_2}) \,f(\Ga)
\ee
and we have to deal only with 12 terms, rather than~16.

The counterterms for the subdiagrams $\ga_1$, $\ga_2$,
$\ga_1 \uplus \ga_2$ can be symbolically written as
\bea
C(\ga_1) &=& - \<\Dl^2(y - x_2), \aux_{\ga_1}(y - x_2)>_y\ \dl(x_1-x_2),
\nn \\
C(\ga_2) &=& - \<\Dl^2(z - x_4), \aux_{\ga_2}(z - x_4)>_z\ \dl(x_3-x_4),
\eea
and according to the choice $\aux_{\ga_1\uplus\ga_2} =
\aux_{\ga_1}\aux_{\ga_2}$ 
\be
C(\ga_1\uplus\ga_2) = \<\Dl^2(y - x_2), \aux_{\ga_1}(y - x_2)>_y
\<\Dl^2(z - x_4), \aux_{\ga_2}(z - x_4)>_z \dl(x_1-x_2) \dl(x_3-x_4),
\ee
and thus
\be
C(\ga_1\uplus\ga_2) = C(\ga_1)C(\ga_2).
\ee
Note that this is \textit{not} a definition: it is the result of a
calculation in which the recursive definition has been exclusively
used.

The previous example was simple in the sense that all divergences
encountered were logarithmic. An example with quadratic
(sub)divergences is given by the diagram (c) in Figure~\ref{fig6}. One
generalizes the factorization results of the examples in the following
lemma.

\begin{lem}
\label{homom}The counterterm map $C$ verifies
$C(\ga_1 \uplus\cdots\uplus \ga_n) = C(\ga_1) \dots C(\ga_n)$.
\end{lem}

\noindent{\bf Proof}.
We argue by induction; the previous example and example (c) in Section~7
show how to start it. We have:
\be
(-S_\ga\bar R_\ga)\, f(\ga) =
-S_\ga \biggl(\; \sum_{\emptyset\subseteq\tilde\ga\subsetneq\ga}
\bigl[ -S_{\tilde\ga_1}\bar R_{\tilde\ga_1} \bigr]
\cdots
\bigl[-S_{\tilde\ga_n}\bar R_{\tilde\ga_n}\bigr] \biggr) f(\ga).
\ee
In order to obtain and interpret this formula, proceed as follows: use the definition of
$\bar R_\ga$ in terms of the subgraphs $\tilde\ga\subset\ga$. 
Now, by the induction
hypothesis, the lemma is true for the $\tilde\ga$.
We apportion $\tilde\ga$ according to the pieces of $\ga$.
A connected piece of a given
$\tilde\ga$ can belong to one of the $\ga_1\cdots\ga_n$ only; but
there could be 
several connected pieces of $\tilde\ga$ in any of the $\ga_1\cdots\ga_n$,
say in $\ga_i$. Using
the induction hypothesis (read in reverse) we treat them as a unity; call it $\tilde\ga_i$. Finally,
if there is no connected piece of $\tilde\ga$ in, say, $\ga_k$, for $k\in\{1\cdots n\}$,
we pretend that there is the piece $\emptyset=\tilde\ga_k\subset\ga_k$.

The previous formula then we rewrite as 
\be
-S_\ga \biggl(
\biggl[\; \sum_{\emptyset\subseteq\tilde\ga_1\subseteq\ga_1}
(-S_{\tilde\ga_1}\bar R_{\tilde\ga_1})\biggr]
\cdots
\biggl[\; \sum_{\emptyset\subseteq\tilde\ga_n\subseteq\ga_n}
(-S_{\tilde\ga_n}\bar R_{\tilde\ga_n})\biggr]
-(-S_{\ga_1}\bar R_{\ga_1})\cdots(-S_{\ga_n}\bar R_{\ga_n})\biggr) f(\ga),
\ee
by adding and subtracting the last term. Now, by the same definition:
\be
\sum_{\emptyset\subseteq\tilde\ga_i\subseteq\ga_i}
(-S_{\tilde\ga_i}\bar R_{\tilde\ga_i}) f(\ga_i) =
(1-S_{\ga_i})\bar R_{\ga_i} f(\ga_i).
\ee
Therefore, the expression is further transformed into
\be
-S_\ga \biggl(
\bigl[ (1-S_{\ga_1})\bar R_{\ga_1}\bigr]
\cdots
\bigl[(1-S_{\ga_n})\bar R_{\ga_n}\bigr]
-(-S_{\ga_1}\bar R_{\ga_1})\dots(-S_{\ga_n}\bar R_{\ga_n})\biggr) f(\ga).
\ee
Because the degree of divergence of $\ga$ is the sum of the degree of
divergence of its connected parts, the first term vanishes under the
factorization
rule for the auxiliary functions (a slight generalization of Lemma 4.3).
Furthermore, $S_{\ga}S_{\ga_1}\cdots S_{\ga_n}=S_{\ga_1}\cdots S_{\ga_n}$.
$\square$

The combinatorics of this proof is identical to the one in the parallel
proof for the BPHZ formalism~\cite{Manoukian}. It is well known, and fairly
clear from the
examples in this section and Section~7, that the recursive formula for
renormalization, for a proper connected diagram $\Ga$, gives rise to
the following nonrecursive formula:
\be
R_\Ga f(\Ga)
 = \Bigl[1 + \sum_\E\prod_{\ga\in\E}(-S_\ga)\Bigr]\,f(\Ga),
\ee
where one sums over all nonempty sets $\E$ whose elements $\ga$ are
proper, divergent, not necessarily connected subdiagrams made of
subgraphs of $\Ga$, that may be $\Ga$ itself, and
$\ga_1,\ga_2 \in \E$ implies that $\ga_1 \subsetneq \ga_2$ or
vice~versa. If $\ga_1 \subsetneq \ga_2$, then the order of the
subtractions is as $\cdots S_{\ga_2}\cdots S_{\ga_1}\cdots$. It is
also rather immediate here, as a consequence of the lemma, that the
previous formula can be rewritten as
\be
R_\Ga f(\Ga)
 = \Bigl[1 + \sum_\F\prod_{\ga\in\F}(-S_\ga)\Bigr]\,f(\Ga),
\ee
with $\F$ now denoting nonempty sets whose elements $\ga$ are proper,
divergent, and \textit{connected} subgraphs of $\Ga$, that may be
$\Ga$ itself, and $\ga_1,\ga_2 \in \F$ implies that either
$\ga_1 \subsetneq \ga_2$ or $\ga_2 \subsetneq \ga_1$ \textit{or} that
$\ga_1,\ga_2$ are disjoint; the order of subtractions is as before.

In other words, we now sum over forests in the sense of Zimmermann,
the difference with~\cite{Zimmermann} being that, since we are in real
space, we sum over subgraphs instead of over subdiagrams. That the
forest formula in this sense holds in the context of the
Epstein--Glaser renormalization must be known to the experts, but we
were unable to find an argument for that in the literature.

For theoretical purposes, there is some advantage in the forest
formula, e.g., for the notorious QED diagram mooted by Kreimer
in~\cite{DirkFirst} there are 32 $\F$-type sets, whereas there are 68
$\E$-type sets. For practical purposes, one of course always uses the
recursive formulae \`a la Bogoliubov--Epstein--Glaser.

(The more trivial cases of renormalization are easily handled. When
$\Ga$ with $E > 4$ is not primitive, we set
\be
R_\Ga f(\Ga) = \bar R_\Ga f(\Ga).
\ee
If $\Ga$ is primitive, then simply
\be
R_\Ga f(\Ga) = W_\Ga f(\Ga),
\lbl{Rprimit}
\ee
i.e., $\bar R_\Ga = 1$. Finally, we can let in fully convergent
diagrams by decreeing that, if $f(\Ga)$ is superficially convergent
and has no subdivergence, simply
\be
R_\Ga f(\Ga) = f(\Ga).)
\ee

\section{A Hopf algebra of graphs for the $\phi^4_4$ theory}

\indent

Now we are all set to prove the compatibility of CPT with the
essential kernel of the Hopf algebra picture of Connes and Kreimer. We
start with a preliminary result. If
$\ga' \subseteq \ga \subseteq \Ga$, then $\ga/\ga'$ is naturally
interpreted as a subdiagram of $\Ga/\ga'$ and there is the following
obvious

\begin{lem}
\label{lm:gagaga}
If $\ga' \subseteq \ga \subseteq \Ga$, then
$(\Ga/\ga')/(\ga/\ga') \simeq \Ga/\ga$.
\end{lem}

A point not emphasized in~\cite{CoKrI} is that the definition of the
Hopf algebra of Feynman graphs associated to a given quantum field
theory can be made in slightly different ways, depending on purpose.
In distinction to reference~\cite{CoKrI}, we allow for improper
diagrams living in our Hopf algebra. Namely, connected graphs will be
considered as the generators instead of only proper ones as in
\cite{CoKrI}. (In fact, it is feasible to allow for \textit{all}
Feynman graphs appearing in the expansion of $\Sb$.) Consideration of
improper diagrams is called for in the Epstein--Glaser formalism, that
leads naturally to connected Green functions and not to 1-particle
irreducible ones.

That said, the algebra $\H$ is defined as the polynomial (hence
commutative) algebra generated by the empty set $\emptyset$ and the
connected Feynman graphs that are (superficially) divergent and/or
have (superficially) divergent subdiagrams, with set union as the
product operation (hence $\emptyset$ is the unit element of~$\H$).

A telling operation in a Hopf algebra $\H$ is often the coproduct
$\trl: \H \to \H \ox \H$; as it is to be a homomorphism of the algebra
structure, we need only define it on connected diagrams. By
definition, the coproduct of $\Ga$, a graph in $\phi^4_4$ theory, is
given by
\be
\trl \Ga\,
 = \sum_{\emptyset\subseteq\ga\subseteq\Ga} \ga \ox \Ga/\ga\,.
\lbl{coproduct}
\ee
The sum is over all divergent, proper, not necessarily connected
subdiagrams of $\Ga$ that are subgraphs or products of subgraphs, such
that \textit{each piece is divergent}, including (and then with the
possible exception of, as $\Ga$ need not be divergent nor proper) the
empty set and $\Ga$ itself ---as the formula makes explicit. If
$\ga = \emptyset$, we write $\ga = 1$ and $\Ga/\ga = \Ga$; if
$\ga = \Ga$, we write $\Ga/\ga = 1$. Note that a nonempty $\Ga/\ga$
will be proper iff $\Ga$ is ---the situation considered
in~\cite{CoKrI}.

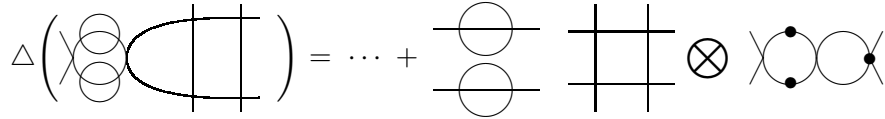
\begin{figure} [!hbp]

\vskip 5mm

$$
\trl\Biggl(\ \,
\parbox{27mm}{\begin{picture}(20,10)
\put(0,5){\line(-1,2){5}}
\put(0,5){\line(-1,-2){5}}
\put(10,5){\circle{20}}
\put(10,14){\circle{15}}
\put(10,-4){\circle{15}}
\put(20,5){\qbezier(0,0)(0,17)(50,15)}
\put(20,5){\qbezier(0,0)(0,-17)(50,-15)}
\put(63,5){\qbezier(0,20)(0,0)(0,-20)}
\put(45,5){\qbezier(0,20)(0,0)(0,-20)}
\end{picture}}
\Biggl) \ = \;\cdots\;+\;
\parbox{18mm}{\begin{picture}(20,10)
\put(0,16){\line(1,0){10}}
\put(20,16){\circle{20}}
\put(10,16){\line(1,0){30}}
\put(0,-7){\line(1,0){10}}
\put(20,-7){\circle{20}}
\put(10,-7){\line(1,0){30}}
\end{picture}}
\parbox{15mm}{\begin{picture}(20,10)
\put(0,15){\line(1,0){40}}
\put(0,-5){\line(1,0){40}}
\put(10,5){\qbezier(0,20)(0,0)(0,-20)}
\put(30,5){\qbezier(0,20)(0,0)(0,-20)}
\end{picture}}
\bx\ \
\parbox{15mm}{\begin{picture}(20,10)
\put(0,5){\line(-1,2){5}}
\put(0,5){\line(-1,-2){5}}
\put(10,5){\circle{20}}
\put(30.2,5){\circle{20}}
\put(40.2,5){\line(1,2){5}}
\put(40.2,5){\line(1,-2){5}}
\put(7.5,12){$\bullet$}
\put(7.5,-7){$\bullet$}
\put(37.5,2){$\bullet$}
\end{picture}}
$$

\caption{A vanishing contribution of a logarithmically divergent
disconnected subdiagram.}\label{fig5}
\end{figure}

Note that the definition of the Hopf algebra coproduct excludes terms
like the one indicated in Figure~\ref{fig5}, that would give anyway zero
contribution because of Lemma~\ref{homom}. This also guarantees that
the graphs $\Ga$ and $\Ga/\ga$ have the same external structure.

If there are no nontrivial subdiagrams in $\Ga$, then this graph will
be primitive both in the sense of quantum field and Hopf algebra
theory.

In Figure~\ref{exples} there are some examples.

\begin{figure} [!hbp]

\vskip 7mm

%
1. On primitive diagrams
\\[4mm]
$
\trl \blp\, \sunset \brp
\ =\ \1 \bx  \sunset +\, \sunset \bx \1 $
\hfill{\footnotesize ``setting sun''}
\\[7mm]
%
$
\trl\blp \p \brp \ =\ \1 \bx \p + \p \bx \1
$
\hfill{\footnotesize (a non planar diagram)\hcm (a)}
\\[1cm]
2. On proper diagrams
\\[4mm]
%
$
\trl \blp \iice \brp
\ =\ \1 \bx \iice
+ 2 \hs{3} \sweet \bx \sunset \put(-35,0){$\bullet$}
+ \iice \bx \1
$\hfill{\parbox{25mm}{\footnotesize
``double ice cream ${}$\hs{5} in a cup''}}
\\[1cm]
%
$
\trl \blp\ \, \tsweet \brp
\ =\ \1 \bx\ \ \tsweet\ + 2\
\ \ \sweet \bx\ \dsweet
\put(-45,0){$\bullet$} \
+ \ \ \sweet \bx\ \dsweet
\put(-25,0){$\bullet$} \
$
\\[5mm]
${} \hskip .7cm
+ 2 \hs{4} \dsweet\  \bx\  \sweet
\put(-31,0){$\bullet$}
+\ \ \sweet\ \ \sweet \bx\ \sweet
\put(-31,0){$\bullet$} \put(-11,0){$\bullet$}
+\ \ \tsweet\ \bx \1
$
\hfill{\footnotesize ``triple sweet''}
\\[1cm]
$
\trl \blp\ \ \kg \brp \ =\
\1 \bx\ \kg +\ \ \sweet \bx \pinterIV
\put(-38,0){$\bullet$}
+\ \ \sweet \bx\ \dsweet
\put(-5,0){$\bullet$}
$
\\
${}$\hfill{(b)}
\\
${} \hskip 2.cm
+\ \ \pinterIV \bx\ \sweet
\put(-11,0){$\bullet$}
+\ \ \sweet\ \ \sweet \bx\ \sweet
\put(-31,0){$\bullet$}
\put(-11,0){$\bullet$}
+\ \ \kg \bx \1
$
\\[1cm]
%
$
\trl \blp \ocateye \brp
\ =\ \1 \bx \ocateye
+\ 2 \ \pinterIV
\bigotimes\ \ \sweet
\put(-31,0){$\bullet$}
+\ \ \sweet \bigotimes\ \ \dsweet
\put(-25,0){$\bullet$}
\ +\ \ocateye\ \bigotimes \1
$
\hfill{\parbox{10mm}{{\footnotesize ``cateye''} ${}$ \hs{4}(c)}}
\\[1cm]
3. On a connected but improper diagram
\\[1cm]
$
\trl \blp \triple \brp
\ =\ \1 \bx \triple
+ 3\ \ \sunset \bx
\parbox{28mm}{\begin{picture}(20,10)
\put(0,10){\line(1,0){80}}
\put(20,10){\circle{20}}
\put(60,10){\circle{20}}
\put(40,-10){\line(0,1){30}}
\put(37.3,-2){$\bullet$}
\end{picture}}
$
\\
${}$ \hfill{\footnotesize ``triple walnut''}
\\[1cm]
${} \qquad + 3\ \
\parbox{15mm}{\begin{picture}(20,10)
\put(0,16){\line(1,0){10}}
\put(20,16){\circle{20}}
\put(10,16){\line(1,0){30}}
\put(0,-7){\line(1,0){10}}
\put(20,-7){\circle{20}}
\put(10,-7){\line(1,0){30}}
\end{picture}}
\ \bx \hskip -5mm
\parbox{23mm}{\begin{picture}(20,10)
\put(20,15){\line(1,0){40}}
\put(27,12.4){$\bullet$}
\put(47,12.4){$\bullet$}
\put(40,-25){\line(0,1){50}}
\put(40,-5){\circle{20}}
\end{picture}}
+\ \ \sunset
\parbox{15mm}{\begin{picture}(20,10)
\put(0,16){\line(1,0){10}}
\put(20,16){\circle{20}}
\put(10,16){\line(1,0){30}}
\put(0,-7){\line(1,0){10}}
\put(20,-7){\circle{20}}
\put(10,-7){\line(1,0){30}}
\end{picture}}
\ \bx \hskip -7mm
\parbox{22mm}{\begin{picture}(20,10)
\put(20,15){\line(1,0){40}}
\put(27,12.4){$\bullet$}
\put(47,12.4){$\bullet$}
\put(40,-10){\line(0,1){30}}
\put(37.3,0){$\bullet$}
\end{picture}}
+\ \triple \bx \1
$
\\[5mm]
\caption{A few examples of coproducts.}\label{exples}
\end{figure}

The counit $\eps: \H \to \C$ is given by $\eps(\Ga)=0$ for all graphs
and $\eps(\emptyset) = 1$. All of the bialgebra axioms are readily
verified, except for coassociativity. The
coproduct in our Hopf algebra of Feynman graphs governs in some sense
the various splittings of a $T$-product at a given order in terms of
lower order $T$-products (see the criterion towards the end of
Section~\ref{ext}), and so coassociativity is of course the soul of
the whole matter. But this is easy enough.

\begin{lem}
The algebra of graphs $\H$ is a Hopf algebra.
\end{lem}

\noindent{\bf Proof}.
We bother only with the proof of coassociativity. We need to show that
$$
((\trl \ox \id) \circ \trl) \Ga = ((\id \ox \trl) \circ \trl) \Ga,
$$
for every connected graph $\Ga$, where $\id$ denotes the identity map
of $\H$ into itself. Using the definition, the left hand side is
written as
$$
\sum_{\emptyset\subseteq\ga'\subseteq\ga\subseteq\Ga}
 \ga' \ox \ga/\ga' \ox \Ga/\ga,
$$
and the right hand side as
$$
\sum_{\emptyset\subseteq\ga'\subseteq\Ga,\;
      \emptyset\subseteq\ga''\subseteq\Ga/\ga'}
 \ga' \ox \ga'' \ox (\Ga/\ga')/\ga''.
$$
We must then show that
$$
\sum_{\ga'\subseteq\ga\subseteq\Ga} \ga/\ga' \ox \Ga/\ga
= \sum_{\emptyset\subseteq\ga''\subseteq\Ga/\ga'}
\ga'' \ox (\Ga/\ga')/\ga''.
$$
Let $\ga' \subseteq \ga \subseteq \Ga$. Then, as already remarked,
$\emptyset \subseteq \ga/\ga' \subseteq \Ga/\ga'$. Reciprocally, to
every $\ga'' \subseteq \Ga/\ga'$ corresponds a $\ga$ such that
$\ga' \subseteq \ga \subseteq \Ga$ and $\ga/\ga' = \ga''$. Then
use Lemma~\ref{lm:gagaga}.\quad  $\square$

Note that if $\#(\Ga)$ denotes the number of vertices in $\Ga$, then
$\H$ is graded by $\# - 1$. To be precise, the degree of a generator
$\Ga$ (connected element) is $\#(\Ga) - 1$; the degree of a product is
the sum of the degrees of the factors. This grading is compatible with
the coproduct. For graded bialgebras, it is easy to construct the antipode,
by methods spelled out in~\cite[Ch.~14]{Polaris}, for instance;
and so $\H$ is a Hopf algebra.

The main role for the definition of coproduct in a coalgebra
is that it allows to define a convolution operation~$\star$ for two
linear maps $f,g$ to any algebra, given by the composition
\be
g \star f \; (h) = \sum g(h')f(h'')
\ee
if $\trl h = \sum h' \ox h''$. In particular, for our
(commutative) Hopf algebra definition~\rf{coproduct},
\be
g \star f \; (\Ga) =
 g(\Ga) + f(\Ga) + \sum_{\emptyset\subsetneq\ga\subsetneq\Ga} g(\ga)\,
f(\Ga/\ga),
\ee
and the convolution of two homomorphisms will be a homomorphism.
Now, because of lemma \ref{homom} in the previous section, the linear map
$C$ is multiplicative. Thus the renormalization formula \rf{renorm}
\be 
R_\Ga f(\Ga) =: R(\Ga) = C(\Ga) + f(\Ga) +
\sum_{\emptyset\subsetneq\ga\subsetneq\Ga} C(\ga)\, f(\Ga/\ga) 
\ee 
is recast as 
\be R = C \star f, \ee 
dropping the variable tag $\Ga$, as we now may. In summary,
the main result of this paper can be stated as follows.

\begin{Thm}
Bogoliubov's renormalization maps $R$ are (distributional) characters
on the Hopf algebra of graphs~$\H$. They are the Hopf convolution of the
(unrenormalized) Feynman graph character and a counterterm character.
\end{Thm}

This is what opens the door to the application of Lie-theoretical
methods characteristic of (co)\-commutative Hopf algebra theory in the
renormalized theory~\cite{CoKrII,Rusos}. In particular, it is clear
that $R$-maps differing in the choice of subtraction operators are
related by elements of a huge ``renormalization group'' sitting inside
$\G_\H$. In other words, having constructed two solutions for the
scattering operator involving two different $T$-products,  the
relationship between their respective ``coupling constants'' is given
by a local formal diffeomorphism~\cite{PoSt}.

\section{Some further examples of CPT renormalization}
\label{check}

\indent

In this Section, we shall renormalize a few other diagrams in
$\phi^4_4$ theory along the lines prescribed by the Epstein--Glaser
scheme according to the previous sections.

\begin{figure}[!ht]
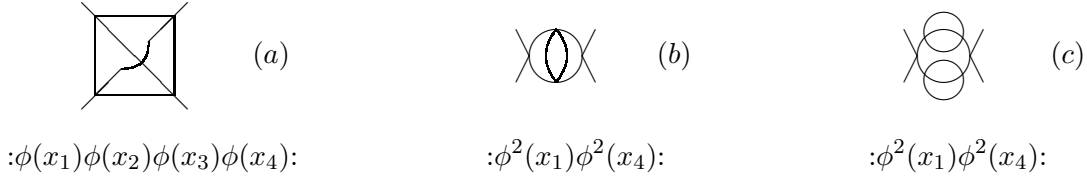


\indent

\begin{eqnarray}
&& \p \qquad (a) \hs{30} \ocateye \qquad (b) \hs{30}
\cafard \qquad (c) \nn\\
\nn\\
&& \hs{-10} \wick{\phi(x_1)\phi(x_2)\phi(x_3)\phi(x_4)} \hs{25}
\wick{\phi^2(x_1)\phi^2(x_4)} \hs{27} \wick{\phi^2(x_1)\phi^2(x_4)}
\nn
\end{eqnarray}
\caption{An interesting primitive graph $(a)$, overlapping logarithmic
subdivergences $(b)$ and quadratic ${}$ \hs{14}
subdivergences, the ``cafard'' $(c)$.}
\label{fig6}
\end{figure}

We shall be concerned with the three following graphs, see
Figure~\ref{fig6}, two of order~4
($(a)$ and $(c)$ of Figure~\ref{exples}) and one of order~6.

All are logarithmically superficially divergent, $\om(\Ga) = 4 - E = 0$,
although the last one has quad\-ratic subdivergences. They come in
$\Sb$ as coefficients of the Wick monomials
$\wick{\phi(x_1)\phi(x_2)\phi(x_3)\phi(x_4)}$ for~$(a)$ and
$\wick{\phi^2(x_1)\phi^2(x_4)}$ for $(b)$ and~$(c)$. We shall not care
about the symmetry factors. In the computation the
criterion~\rf{crit} for the Epstein--Glaser recursive construction
will be applied but
only for partitionings associated to divergent, proper, not
necessarily connected subgraphs.

\subsection*{Graph (a)}

\indent

Numbering the vertices clockwise, one gets the operator
\begin{eqnarray}
\leqa{ \<R_\Ga f(\Ga),\varphi_\Ga> }
\nn\\
&& \propto \ \la^4 \
\< \bigl[\Dl(x_1 - x_2)\Dl(x_1 - x_3) \Dl(x_1 - x_4 )
\Dl(x_2 - x_3) \Dl(x_3 -x_4) \Dl(x_4 -x_2)\bigr]_R,
\varphi_\Ga(x_1,x_2,x_3,x_4)>, \nn
\end{eqnarray}
where $\varphi_\Ga(x_1,x_2,x_3,x_4)$ stands for
$\wick{\phi(x_1)\phi(x_2)\phi(x_3)\phi(x_4)}\ g(x_1)g(x_2)g(x_3)g(x_4)$.
According to \rf{Rprimit} one obtains
\begin{eqnarray}
&&\langle \Dl(x_1 - x_2)\Dl(x_1 - x_3) \Dl(x_1 - x_4 )
\Dl(x_2 - x_3) \Dl(x_3 -x_4) \Dl(x_4 -x_2),
\nn\\
&&\hs{30} \wick{\phi(x_1)\phi(x_2)\phi(x_3)\phi(x_4)}\
g(x_1)g(x_2)g(x_3)g(x_4)
\nn\\
&&\hs{40} -\
\aux_{\Ga}(x_1-x_2,x_2-x_3,x_3-x_4)\wick{\phi^4(x_4)}g^4(x_4) \rangle
\nn
\end{eqnarray}
corresponding to the following pattern of subtraction:
\begin{eqnarray}
f(\Ga) \mapsto (1 - S_{\Ga})\,f(\Ga).
\end{eqnarray}

\subsection*{Graph (b)}

\indent

Numbering the vertices from left to right, this corresponds to the operator
\begin{eqnarray}
\leqa{ \<R_\Ga f(\Ga),\varphi_\Ga> }
\nn\\
&& \propto \ \la^4 \
\< \bigl[\Dl(x_1-x_2)\Dl(x_1-x_3)\Dl^2(x_2-x_3)\Dl(x_2-x_4)
  \Dl(x_3-x_4)\bigr]_R, \varphi_\Ga(x_1,x_2,x_3,x_4)>,
\nn
\end{eqnarray}
where $\varphi_\Ga(x_1,x_2,x_3,x_4)$ stands for
$\wick{\phi^2(x_1)\phi^2(x_4)}\ g(x_1)g(x_2)g(x_3)g(x_4)$. Therefore,
\begin{eqnarray}
\leqa{ \< \bar R_\Ga f(\Ga), W_\Ga \vf_\Ga> =
\langle \bigl[ \Dl(x_1-x_2)\Dl(x_1-x_3)\Dl^2(x_2-x_3)
  \Dl(x_2-x_4)\Dl(x_3-x_4) \bigr]_{\bar R}\, ,}
\nn\\
&& \wick{\phi^2(x_1)\phi^2(x_4)}\ g(x_1)g(x_2)g(x_3)g(x_4)
   - \aux_{\Ga}(x_1-x_2,x_2-x_3,x_3-x_4)\wick{\phi^4(x_4)}
   g^4(x_4) \rangle. \nn
\end{eqnarray}
Taking into account all three logarithmically divergent subgraphs, the
whole ice-cream cones on the left $\ga_1$ and on the right $\ga_2$,
and the central fish $\ga_3 = \ga_1 \cap \ga_2$, which is the
overlap of the former,

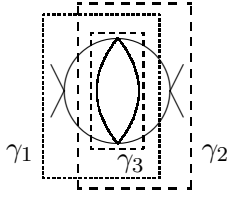
\begin{figure}[!ht]

\vskip 1cm

\begin{center}
\parbox{42mm}{\begin{picture}(20,10)
\put(-8,-28){\dashbox{1}(44,62)} \put(-22,-20){$\ga_1$}
\put(0,5){\line(-1,2){5}}
\put(0,5){\line(-1,-2){5}}
\put(20,5){\circle{40}}
\put(40,5){\line(1,2){5}}
\put(40,5){\line(1,-2){5}}
\put(5,-32){\dashbox{3}(43,70)} \put(52,-20){$\ga_2$}
\put(10,-17){\dashbox{2}(20,44)} \put(20,-24){$\ga_3$}
\qbezier(12,5)(12,15)(20,25)
\qbezier(28,5)(28,15)(20,25)
\qbezier(12,5)(12,-5)(20,-15)
\qbezier(28,5)(28,-5)(20,-15)
\end{picture}}
\end{center}

\vskip 7mm

\caption{Identification of proper subdiagrams with an overlapping
logarithmic divergence.}
\label{fig7}
\end{figure}

\noindent
one successively gets
\begin{eqnarray}
&& \langle \Dl(x_1-x_2)\Dl(x_1-x_3)\Dl^2(x_2-x_3)\Dl(x_2-x_4)
\Dl(x_3-x_4), \wick{\phi^2(x_1)\phi^2(x_4)}\ g(x_1)g(x_2)g(x_3)g(x_4)
\nn \\
&&\hs{30} -\
\aux_{\Ga}(x_1-x_2,x_2-x_3,x_3-x_4)\wick{\phi^4(x_4)}g^4(x_4) \rangle
\nn\\
&& -\ \langle 
\bigl[ \Dl(x_1-x_2)\Dl(x_1-x_3)\Dl^2(x_2-x_3) \bigr]_{\bar R}
\Dl^2(x_3-x_4),
\nn\\
&&\hs{30} \aux_{\ga_1}(x_1-x_2,x_2-x_3)\wick{\phi^2(x_1)\phi^2(x_4)}\
g^3(x_1)g(x_4)
\nn\\
&&\hs{30} - \aux_{\ga_1}(x_1-x_2,x_2-x_3)\aux_{\Ga}(0,0,x_3-x_4)
   \wick{\phi^4(x_4)}g^4(x_4) \rangle
\lbl{ga1} \\
&& -\ \langle
\bigl[ \Dl^2(x_2-x_3)\Dl(x_2-x_4)\Dl(x_3-x_4)\bigr]_{\bar R}
\Dl^2(x_1-x_2),
\nn\\
&&\hs{30} \aux_{\ga_2}(x_2-x_3,x_3-x_4)\wick{\phi^2(x_1)\phi^2(x_4)}\
g(x_1)g^3(x_4)
\nn\\
&&\hs{30} - \aux_{\ga_2}(x_2-x_3,x_3-x_4)\aux_{\Ga}(x_1-x_2,0,0)
\wick{\phi^4(x_4)}g^4(x_4) \rangle
\lbl{ga2}\\
&& -\ \langle \Dl^2(x_1-x_2) \Dl^2(x_2-x_3)\Dl^2(x_3-x_4),
\aux_{\ga_3}(x_2-x_3)\wick{\phi^2(x_1)\phi^2(x_4)}\
g(x_1)g^2(x_2)g(x_4)
\nn\\
&&\hs{30} -\ \aux_{\ga_3}(x_2-x_3)\aux_{\Ga}(x_1-x_2,0,x_3-x_4)
\wick{\phi^4(x_4)}g^4(x_4) \rangle.
\nn
\end{eqnarray}
Incorporating the renormalization of the subgraphs $\ga_1$ and $\ga_2$
(previously obtained at the lower order~3) yields the replacement of
the pairings \rf{ga1} and \rf{ga2} respectively by
\begin{eqnarray}
&& \rf{ga1}
 = - \langle \Dl(x_1-x_2)\Dl(x_1-x_3)\Dl^2(x_2-x_3)\Dl^2(x_3-x_4),
\nn\\
&&\hs{30} \aux_{\ga_1}(x_1-x_2,x_2-x_3)\wick{\phi^2(x_1)\phi^2(x_4)}\
g^3(x_1)g(x_4)
\nn\\
&&\hs{30} - \aux_{\ga_1}(x_1-x_2,x_2-x_3)\aux_{\Ga}(0,0,x_3-x_4)
\wick{\phi^4(x_4)}g^4(x_4) \rangle
\nn \\
&& + \langle \Dl^2(x_1-x_2) \Dl^2(x_2-x_3)\Dl^2(x_3-x_4),
\aux_{\ga_3}(x_2-x_3)\aux_{\ga_1}(x_1-x_2,0)
\wick{\phi^2(x_1)\phi^2(x_4)}\ g^3(x_1)g(x_4)
\nn\\
&&\hs{30} - \aux_{\ga_3}(x_2-x_3)\aux_{\ga_1}(x_1-x_2,x_2-x_3)
\aux_{\Ga}(0,0,x_3-x_4) \wick{\phi^4(x_4)}g^4(x_4) \rangle
\nn\\[2mm]
&&\rf{ga2}
= - \langle \Dl^2(x_1-x_2)\Dl^2(x_2-x_3)\Dl(x_2-x_4)\Dl(x_3-x_4),
\nn\\
&&\hs{30} \aux_{\ga_2}(x_2-x_3,x_3-x_4)\wick{\phi^2(x_1)\phi^2(x_4)}\
g(x_1)g^3(x_4)
\nn\\
&&\hs{30} - \aux_{\ga_2}(x_2-x_3,x_3-x_4)\aux_{\Ga}(x_1-x_2,0,0)
\wick{\phi^4(x_4)}g^4(x_4) \rangle
\nn\\
&& + \langle \Dl^2(x_1-x_2) \Dl^2(x_2-x_3)\Dl^2(x_3-x_4),
\aux_{\ga_3}(x_2-x_3)\aux_{\ga_2}(0,x_3-x_4)
\wick{\phi^2(x_1)\phi^2(x_4)}\  g(x_1)g^3(x_4)
\nn\\
&&\hs{30} - \aux_{\ga_3}(x_2-x_3)\aux_{\ga_2}(0,x_3-x_4)
\aux_{\Ga}(x_1-x_2,0,0) \wick{\phi^4(x_4)}g^4(x_4) \rangle.
\nn
\end{eqnarray}
This graph has the following pattern of subtraction:
\begin{eqnarray}
f(\Ga) \mapsto (1 - S_{\Ga})[ 1 - S_{\ga_1}(1 - S_{\ga_3})
 - S_{\ga_2}(1 - S_{\ga_3}) - S_{\ga_3} ] \,f(\Ga)\ .
\end{eqnarray}

\subsection*{Graph (c)}

\indent

Among all the operators contributing to the sixth order one considers
the one associated to the ``cafard'' graph.  It is mathematically convenient,
for this kind of diagrams, to assume that all auxiliary functions used
have vanishing derivatives at the coincidence points up to the highest
order of divergence encountered in the diagram (failure to do so can
always be corrected by an appropriate choice of finite counterterms).
Also as in~\cite{Pinter}, it is recalled that when some ``coupling
constants'' $g$ become derivated according to the $W$ projection, the
corresponding terms vanish in the adiabatic limit (in the strong sense
in massive $\phi^4_4$ theory). Thus such terms can be
discarded. Numbering the six vertices clockwise, one has the operator
\begin{eqnarray}
\<R_\Ga f(\Ga),\varphi_\Ga> \hspace{-5mm}
&& \propto \ \la^6 \ \langle \bigl[\Dl(x_1 - x_2)\Dl^3(x_2 - x_3)
\Dl(x_3 - x_4)\Dl(x_4 -x_5)\Dl^3(x_5 - x_6)\Dl(x_6 - x_1) \bigr]_R,
\nn\\
& &\hs{40} \varphi_\Ga(x_1,x_2,x_3,x_4,x_5,x_6) \rangle,
\lbl{o6}
\end{eqnarray}
where $\varphi_\Ga(x_1,x_2,x_3,x_4,x_5,x_6)$ stands for
$\wick{\phi^2(x_1)\phi^2(x_4)}g(x_1)g(x_2)g(x_3)g(x_4)g(x_5)g(x_6)$.
There are three graphs to be considered, namely, both the top and
bottom sunsets $\ga_1$ and $\ga_2$ respectively which are quadratically
divergent, $\om(\ga_i) = 2$, $i = 1,2$ and the union of the two
$\ga_1 \uplus \ga_2$ with power counting index
$\om(\ga_1\uplus\ga_2) = \om(\ga_1) + \om(\ga_2) = 4$ according to the
coproduct:

\begin{eqnarray}
\trl \blp\ \ \cafard  \brp
= \1 \ox\ \ \cafard + 2\ \sunset \ox\
\parbox{10mm}{\begin{picture}(20,10)
\put(0,5){\line(-1,2){5}}
\put(0,5){\line(-1,-2){5}}
\put(10,5){\circle{20}} \put(10,14){\circle{15}}
\put(7.5,-7){$\bullet$}
\put(20,5){\line(1,2){5}}
\put(20,5){\line(1,-2){5}}
\end{picture}}
+\
\parbox{15mm}{\begin{picture}(20,10)
\put(0,16){\line(1,0){10}}
\put(20,16){\circle{20}}
\put(10,16){\line(1,0){30}}
\put(0,-7){\line(1,0){10}}
\put(20,-7){\circle{20}}
\put(10,-7){\line(1,0){30}}
\end{picture}}
\ \ox\ \
\parbox{10mm}{\begin{picture}(20,10)
\put(0,5){\line(-1,2){5}}
\put(0,5){\line(-1,-2){5}}
\put(10,5){\circle{20}}
\put(7.5,12){$\bullet$}
\put(7.5,-7){$\bullet$}
\put(20,5){\line(1,2){5}}
\put(20,5){\line(1,-2){5}}
\end{picture}}
+\ \ \cafard \ox \1\ .\nn
\end{eqnarray}

\vs{3}

\noindent
So \rf{o6} reads
\begin{eqnarray}
&& \hs{-15}\< \bar R_\Ga f(\Ga), W_\Ga \vf_\Ga>
 = \langle \bigl[ \Dl(x_1-x_2)\Dl^3(x_2-x_3)\Dl(x_3-x_4)\Dl(x_4-x_5)
\Dl^3(x_5-x_6)\Dl(x_6-x_1)\bigr]_{\bar R}\, , \nn\\[2mm]
&& \wick{\phi^2(x_1)\phi^2(x_4)}\ g(x_1)g(x_2)g(x_3)g(x_4)g(x_5)g(x_6)
\nn\\[2mm]
&& \hs{20}- \aux_{\Ga}(x_1-x_2,x_2-x_3,x_3-x_4,x_4-x_5,x_5-x_6)
\wick{\phi^4(x_4)}g^6(x_4) \rangle.
\nn
\end{eqnarray}
Then we perform the renormalization by the use of previously
renormalized graphs according to their respective $W$ projection. 
For notational convenience, the pairing will be
done with the projection $W_\Ga \vf_\Ga$ and $\Dl^{(a)}$ below will
denote the derivative of order $|a|$ of the propagator
with respect to the multi-quadri-index $a$, so the previous expression
becomes
\begin{eqnarray}
\leqa{ \langle \Dl(x_1-x_2)\Dl^3(x_2-x_3)\Dl(x_3-x_4)\Dl(x_4-x_5)
\Dl^3(x_5-x_6)\Dl(x_6-x_1)\, , W_\Ga \vf_\Ga \rangle }
\nn\\[1mm]
&& - \sum_{|a|=0}^2\
\sum_{\shortstack{\ssz$|b|,|c|=0$ \\ \ssz$b+c=a$}}^{|a|}
\biggl< \Dl^{(b)}(x_3-x_1)\Dl^3(x_2-x_3)\Dl(x_3-x_4)\Dl(x_4-x_5)
\Dl^3(x_5-x_6)\Dl(x_1-x_6)\, ,
\nn\\[-4mm]
&&\hs{50} \aux_{\ga_1} (x_2-x_3)\, \frac{(x_2-x_3)^a}{b!\,c!}
\left(\frac{\del^{|c|} W_\Ga \vf_\Ga}{\del x_2^c}
\right)\Bigr|_{x_2=x_3} \biggr>
\nn\\
&& -  \sum_{|a|=0}^2\
\sum_{\shortstack{\ssz$|b|,|c|=0$ \\ \ssz$b+c = a$}}^{|a|}
\biggl< \Dl(x_1-x_2)\Dl^3(x_2-x_3)\Dl(x_3-x_4)\Dl^{(b)}(x_6-x_4)
\Dl^3(x_5-x_6)\Dl(x_1-x_6)\, ,
\nn\\[-4mm]
&&\hs{50} \aux_{\ga_2} (x_5-x_6)\, \frac{(x_5-x_6)^a}{b!\,c!}
\left(\frac{\del^{|c|} W_\Ga \vf_\Ga}{\del x_5^c}
\right)\Bigr|_{x_5=x_6} \biggr>
\nn\\[3mm]
&& - \sum_{\shortstack{\ssz$|a|=0$\\ \ssz$ a=a_1+a_2$\\
\ssz$|a_1|,|a_2|\leq 2$}}^4 \ \
\sum_{\shortstack{\ssz$|b|=|b_1|+|b_2|=0$ \\ \ssz$|c|=|c_1|+|c_2|=0$\\
\ssz$b+c=a$}}^{|a|}
\biggl< \bigl[\Dl^3(x_2-x_3)\Dl^3(x_5-x_6)\bigr]_{\bar R}
\nn\\[-8mm]
&& \hs{70} \Dl^{(b_1)}(x_3-x_1)\Dl(x_3-x_4)\Dl^{(b_2)}(x_6-x_4)\Dl(x_6-x_1)\, ,
\nn\\[5mm]
&& \hs{-5}
\aux_{\ga_1\uplus\ga_2} (x_2-x_3,x_5-x_6)\,
\frac{(x_2-x_3)^{a_1}(x_5-x_6)^{a_2}}{b!\,c!}
\left(\frac{\del^{|c|} W_\Ga \vf_\Ga}
{\del x_2^{c_1}\del x_5^{c_2}} \right)
\Bigr|_{\shortstack{\ssz$x_2=x_3$\\ \ssz$x_5=x_6$}}\ \biggr>
\lbl{ga12}
\end{eqnarray}
In the last pairing \rf{ga12}, the summation on the multi-quadri-index
$a = (a_1,a_2)$ is restricted to the condition $|a_1|,|a_2| \leq 2$ due
to the fact that $\ga_1 \uplus \ga_2$ is a disconnected graph. For instance,
pairing the tensor product distribution 
$\bigl[ \Dl^3(x_2-x_3) \Dl^3(x_5-x_6)\bigr]_{\bar R}$ with a test
function $\vf(x_2-x_3)$ yields a distribution in $x_5-x_6$ of singular
order~2. Then finally \rf{ga12} is worked
out by using the renormalized quadratically divergent subgraphs, as
follows:
\begin{eqnarray}
\leqa{ \rf{ga12}
=\ - \hs{-3} \sum_{\shortstack{\ssz$|a|=0$\\ \ssz$ a=a_1+a_2$\\
\ssz$|a_1|,|a_2|\leq 2$}}^4 \ \
\sum_{\shortstack{\ssz$|b|=|b_1|+|b_2|=0$\\ \ssz$|c|=|c_1|+|c_2|=0$\\
\ssz $b+c=a$}}^{|a|} 
\biggl< \Dl^3(x_2-x_3)\Dl^3(x_5-x_6) }
\nn\\[-8mm]
&& \hs{70} \Dl^{(b_1)}(x_3-x_1)\Dl(x_3-x_4)\Dl^{(b_2)}(x_6-x_4)\Dl(x_6-x_1)\, ,
\nn\\[5mm]
&&  \aux_{\ga_1\uplus\ga_2} (x_2-x_3,x_5-x_6)\,
\frac{(x_2-x_3)^{a_1}(x_5-x_6)^{a_2}}{b!\,c!}
\left(\frac{\del^{|c|} W_\Ga \vf_\Ga}
{\del x_2^{c_1}\del x_5^{c_2}} \right)
\Bigr|_{\shortstack{\ssz$x_2=x_3$\\ \ssz$x_5=x_6$}}
\nn\\[5mm]
&& - \aux_{\ga_1}(x_2-x_3) \sum_{|a'|=0}^2 \
\frac{(x_2-x_3)^{a'} (x_5-x_6)^{a_2}}{b!\,c!\,a'!}
\underbrace{\left[\frac{\del^{|a'|}
[\aux_{\ga_1\uplus\ga_2}(x_2-x_3,x_5-x_6) (x_2-x_3)^{a_1}]}
{\del x_2^{a'}} \right]\Bigr|_{x_2=x_3}}_{\mbox
 {\nms $a_1!\,\aux_{\ga_1\uplus\ga_2}(0,x_5-x_6)\,\dl^{a_1}_{a'}$}}
\nn\\[3mm]
&& \hs{50}
\left(\frac{\del^{|c|} W_\Ga \vf_\Ga}
{\del x_2^{c_1}\del x_5^{c_2}} \right)
\Bigr|_{\shortstack{\ssz$x_2=x_3$\\ \ssz $x_5=x_6$}}
\nn\\[5mm]
&& - \aux_{\ga_2} (x_5-x_6) \sum_{|a'|=0}^2 \
\frac{(x_5-x_6)^{a'} (x_2-x_3)^{a_1}}{b!\,c!\,a'!}
\underbrace{\left[\frac{\del^{|a'|}[
 \aux_{\ga_1\uplus\ga_2}(x_2-x_3,x_5-x_6)(x_5-x_6)^{a_2}]}
 {\del x_5^{a'}} \right]\Bigr|_{x_5=x_6} }_{\mbox{\nms
$a_2!\,\aux_{\ga_1\uplus\ga_2}(x_2-x_3,0)\,\dl^{a_2}_{a'}$} }
\nn\\[3mm]
&& \hs{50}
\left(\frac{\del^{|c|} W_\Ga \vf_\Ga}
{\del x_2^{c_1}\del x_5^{c_2}} \right)
\Bigr|_{\shortstack{\ssz$x_2=x_3$\\ \ssz$x_5=x_6$}}\ \biggr>.
\nn
\end{eqnarray}
Remembering the judicious choice
$\aux_{\ga_1\uplus\ga_2} = \aux_{\ga_1}\aux_{\ga_2}$, the first term
is compensated by one of the two last terms, so that for \rf{o6} one
ends with
\begin{eqnarray}
\leqa{ \langle \Dl(x_1-x_2)\Dl^3(x_2-x_3)\Dl(x_3-x_4)\Dl(x_4-x_5)
\Dl^3(x_5-x_6)\Dl(x_6-x_1)\, , W_\Ga \vf_\Ga \rangle }
\nn\\[3mm]
&& - \sum_{|a|=0}^2\
\sum_{\shortstack{\ssz$|b|,|c|=0$ \\ \ssz$b+c=a$}}^{|a|}
\biggl< \Dl(x_1-x_2)\Dl^3(x_2-x_3)\Dl(x_3-x_4)\Dl(x_4-x_5)
\Dl^3(x_5-x_6)\Dl(x_1-x_6)\, ,
\nn\\[-4mm]
&&\hs{50} \aux_{\ga_1} (x_2-x_3)\, \frac{(x_2-x_3)^a}{b!\,c!}
\left(\frac{\del^{|c|} W_\Ga \vf_\Ga}{\del x_2^c}
\right)\Bigr|_{x_2=x_3} \biggr>
\nn\\
&& - \sum_{|a|=0}^2\
\sum_{\shortstack{\ssz$|b|,|c|=0$ \\ \ssz$b+c = a$}}^{|a|}
\biggl< \Dl(x_1-x_2)\Dl^3(x_2-x_3)\Dl(x_3-x_4)\Dl^{(b)}(x_6-x_4)
\Dl^3(x_5-x_6)\Dl(x_1-x_6)\, ,
\nn\\[-4mm]
&&\hs{50} \aux_{\ga_2} (x_5-x_6)\, \frac{(x_5-x_6)^a}{b!\,c!}
\left(\frac{\del^{|c|} W_\Ga \vf_\Ga}{\del x_5^c}
\right)\Bigr|_{x_5=x_6} \biggr>
\nn\\[3mm]
&& + \sum_{\shortstack{\ssz$|a|=0$\\ \ssz$a=a_1+a_2$\\
\ssz$|a_1|,|a_2|\leq 2$}}^4 \ \
\sum_{\shortstack{\ssz$|b|=|b_1|+|b_2|=0$\\ \ssz$|c|=|c_1|+|c_2|=0$\\
\ssz$b+c=a$}}^{|a|} 
\biggl< \Dl^3(x_2-x_3)\Dl^3(x_5-x_6)
\nn\\[-8mm]
&& \hs{70} \Dl^{(b_1)}(x_3-x_1)\Dl(x_3-x_4)\Dl^{(b_2)}(x_6-x_4)\Dl(x_6-x_1)\, ,
\nn\\[5mm]
&& \aux_{\ga_1}(x_2-x_3) \aux_{\ga_2}(x_5-x_6)\,
\frac{(x_2-x_3)^{a_1}(x_5-x_6)^{a_2}}{b!\,c!}
\left(\frac{\del^{|c|} W_\Ga \vf_\Ga}
{\del x_2^{c_1}\del x_5^{c_2}}\right)
\Bigr|_{\shortstack{\ssz$x_2=x_3$\\ \ssz$x_5=x_6$}}\ \biggr> \, ,
\nn
\end{eqnarray}
a pairing which gives rise to the following subtraction pattern:
\begin{eqnarray}
f(\Ga) \mapsto (1 - S_{\Ga})[ 1 - S_{\ga_1} - S_{\ga_2}
 + S_{\ga_1}S_{\ga_2} ] \,f(\Ga)\ .
\end{eqnarray}
Accordingly, counterterms all come from connected diagrams, and we see
immediately that
\be
C(\ga_1) = - \sum_{|a_1|=0}^2 \biggl<\Dl^3(y-x_3),\aux_{\ga_1}(y-x_3)
\frac{(y-x_3)^{a_1}}{a_1!} \biggr>_{\!\!y} \, (-1)^{|a_1|}
\dl^{(a_1)}(x_2 - x_3), 
\ee
and
\be
C(\ga_2) = - \sum_{|a_2|=0}^2 \biggl< \Dl^3(z-x_6),\aux_{\ga_2}(z-x_6)
\frac{(z-x_6)^{a_2}}{a_2!} \biggr>_{\!\!z} \, (-1)^{|a_2|}
\dl^{(a_2)}(x_5 - x_6).
\ee
It is readily seen that
$C(\ga_1\uplus\ga_2) = C(\ga_1)C(\ga_2)$ since
\begin{eqnarray*}
&& \hs{-5} C(\ga_1\uplus\ga_2) = \hs{-5} \sum_{\shortstack{\ssz$|a|=0$\\
\ssz$a=a_1+a_2$\\\ssz$|a_1|,|a_2|\leq 2$}}^4
\biggl<\Dl^3(y-x_3),\aux_{\ga_1}(y-x_3)
\frac{(y-x_3)^{a_1}}{a_1!} \biggr>_{\!\!y} \\ 
&& \hs{35} \biggl< \Dl^3(z-x_6),\aux_{\ga_2}(z-x_6)
\frac{(z-x_6)^{a_2}}{a_2!} \biggr>_{\!\!z}
(-1)^{|a|} \dl^{(a_1)}(x_2 - x_3)\, \dl^{(a_2)}(x_5 - x_6).
\end{eqnarray*}

\section{Conclusions}

\indent

Despite its elegance and accuracy, the CPT scheme of Epstein and
Glaser was thought not to yield explicit formulae of actual
computational value. Nevertheless, the Z\"urich School kept the
flame; we want to cite here, apart from the textbook by Scharf~\cite{Scharf}
and~\cite{Dut}, the outstanding paper by Hurth and Skenderis~\cite{QNoether}
on CPT for gauge theories, among its more recent productions. Also the
Moscow INR School for a long while made systematic use of
distributional methods; in this respect~\cite{KuzTV}, for instance,
makes fascinating reading. The computational links highlighted by
people from the Hamburg School~\cite{Prange,BrFr99,Pinter} have
contributed, to change the picture. It is important to
realize that CPT suffers none of the limitations of dimensional
regularization (for a nonstandard example, see~\cite{CollinsT}) and
that it has naturally associated to it regularization schemes that
comprise the MS scheme~\cite{KuzT}. On the other hand, the Hopf algebra
approach points to a continent of symmetry lying beyond the sea mists of
the present formalism of quantum field theory~\cite{Rusos}; the versatility
of CPT is surely an important asset for unveiling that continent of symmetry.

The change of renormalization scheme leads us to the use of subgraphs, rather
than more general subdiagrams;
other differences between our method of proof and the treatment in
\cite{CoKrI} by Connes and Kreimer are wholly superficial: for instance, the
fact that they had a particular regularization at their disposal,
allowed them to formulate the basic argument in a more algebraic way,
whereas in our case it boils down to concrete analytical properties of
the maps $W$ of Epstein and Glaser. Another minor difference concerns
the place accorded here to improper diagrams. Let us note, nevertheless,
that a congeries of improper diagrams realize the Connes--Moscovici Lie
algebra~\cite{CM} relation $[Z_m, Z_n] = (m - n) Z_{m+n}$ inside our
Hopf algebra $\H$~\cite{CoP}. This commutator can be graphically
represented as
\begin{equation}
[ \
\parbox{33mm}{\begin{picture}(20,10)
\put(0,5){\line(1,0){10}}
\put(5,5){ $\underbrace{ \begin{picture}(20,10)
\put(0,0){\line(1,0){5}}
\multiput(7.5,0)(12.5,0){2}{\put(0,0){\circle*{5}}\put(2.5,0){\line(1,0){10}} }
\end{picture} \hs{5} \dots\  
\begin{picture}(20,10) 
\put(0,0){\line(1,0){10}}
\put(12.5,0){\circle*{5}}\put(15,0){\line(1,0){5}}
\end{picture} }_{n\ \mbox{\ssz vertices}}$ } 
\put(80,5){\line(1,0){10}}
\end{picture}}, 
\parbox{33mm}{\begin{picture}(20,10)
\put(0,5){\line(1,0){10}}
\put(5,5){ $\underbrace{ \begin{picture}(20,10)
\put(0,0){\line(1,0){5}}
\multiput(7.5,0)(12.5,0){2}{\put(0,0){\circle*{5}}\put(2.5,0){\line(1,0){10}} }
\end{picture} \hs{5} \dots\  
\begin{picture}(20,10) 
\put(0,0){\line(1,0){10}}
\put(12.5,0){\circle*{5}}\put(15,0){\line(1,0){5}}
\end{picture} }_{m\ \mbox{\ssz vertices}}$ } 
\put(80,5){\line(1,0){10}}
\end{picture}}]\ =\
(m-n)\
\parbox{33mm}{\begin{picture}(20,10)
\put(0,5){\line(1,0){10}}
\put(5,5){ $\underbrace{ \begin{picture}(20,10)
\put(0,0){\line(1,0){5}}
\multiput(7.5,0)(12.5,0){2}{\put(0,0){\circle*{5}}\put(2.5,0){\line(1,0){10}} }
\end{picture} \hs{5} \dots\  
\begin{picture}(20,10) 
\put(0,0){\line(1,0){10}}
\put(12.5,0){\circle*{5}}\put(15,0){\line(1,0){5}}
\end{picture} }_{n+m\ \mbox{\ssz vertices}}$ } 
\put(80,5){\line(1,0){10}}
\end{picture}} \.
\end{equation}

\medskip

``Real life'' theories, i.e., gauge theories, also represent a
challenge for the Hopf algebra approach to the renormalization group.
In real life, some of our arguments may not strictly apply; for
instance, normalization conditions or the convenience of preserving
gauge invariance at some stage of the procedure might demand
``oversubtractions'', which effectively represent an increase of the
order of divergence of some diagrams, propagating itself to
higher-order diagrams. It looks likely, however, that these technical
difficulties will be overcome.

\subsection*{Acknowledgments}

\indent

We are very grateful to Philippe Blanchard, Ricardo Estrada, Steve
Fulling, Rainer H\"aussling, Tom\'{a}\v{s} Kopf,
Thomas Krajewski, Dirk Kreimer, Denis Perrot, Florian Scheck,
Raymond Stora, Oleg Tarasov, Joe V\'arilly and Raimar Wulkenhaar for
discussions and/or reading early drafts of the manuscript.

The first-named author (JMGB) thanks the labs at Mainz, Madrid and
Marseille for their warm hospitality and financial support during the
elaboration of the present work.

\end{document}